\documentclass{aa}     
\usepackage{graphicx}
\usepackage{upgreek} 

\usepackage[varg]{txfonts}
\usepackage{natbib}

\usepackage{hyperref}

\hypersetup{%
  colorlinks=true,
  linkcolor=blue,
  citecolor=blue
}

\newcommand{\OI}{O\,{\sc i}}

\newcommand{\FeII}{Fe\,{\sc ii}}

\newcommand{\CII}{C\,{\sc ii}}
\newcommand{\HII}{H\,{\sc ii}}
\newcommand{\HI}{H\,{\sc i}}

\def\Lsun{{\hbox {L$_{\odot}$}}} 
\def\Msun{{\hbox {M$_{\odot}$}}}

\def\PzeroOi{{[O~{\scriptsize I}]~{\scriptsize ${}^3$P$_0-{}^3$P$_1$}}}
\def\PoneOi{{[O~{\scriptsize I}]~{\scriptsize ${}^3$P$_1-{}^3$P$_2$}}}

\def\32OH{{OH}~{$^{2}\Pi_{3/2}$ $J$=$5/2$-$3/2$}}
\def\12OH{{OH}~{$^{2}\Pi_{1/2}$ $J$=$3/2$-$1/2$}}
\def\Cp{{[C~{\scriptsize II}]~{\scriptsize ${}^2$P$_{3/2}-{}^2$P$_{1/2}$}}}

\begin{document} 

\title{Velocity-resolved [\OI]\,63,145\,$\upmu$m, [\CII]\,158\,$\upmu$m, and OH 
mapping along the Orion BN/KL explosive outflow and irradiated shocks\thanks{Dedicated to the memory of Karl M. Menten (1957–2024), whose support for the GREAT developments and whose interest in Orion \mbox{inspired} these observations.}}

\titlerunning{SOFIA velocity-resolved FIR atomic fine-structure line
mapping along Orion BN/KL outflows} 
\authorrunning{Goicoechea et al.}

 \author{Javier R.\,Goicoechea\inst{1}
          \and
    Rolf G\"{u}sten\inst{2}
    \and
    Benjamin Godard\inst{3,4}  
    \and
    Helmut Wiesemeyer\inst{2}  
    \and
    Ronan Higgins\inst{5}  
    \and\\
    Antoine Gusdorf\inst{4,3}  
    \and
    \mbox{Miriam G. Santa-Maria}\inst{1}  
    \and
    Marion Zannese\inst{1}
    \and
    Karl M. Menten\inst{2}  
    }

\institute{Instituto de F\'{\i}sica Fundamental
     (CSIC). Calle Serrano 121-123, 28006, Madrid, Spain. \email{javier.r.goicoechea@csic.es}
\and
Max-Planck Institut f\"{u}r Radioastronomie, Auf dem H\"{u}gel  69, 53121 Bonn, Germany
\and
Observatoire de Paris, Universit\'e PSL, Sorbonne Universit\'e, LUX, 75014 Paris, France
\and
Ecole Normale Sup\'erieure, ENS, Universit\'e PSL, CNRS, Sorbonne Universit\'e, Universit\'e de Paris, 75005 Paris, France
\and
Physikalisches Institut der Universit\"at zu K\"oln, Z\"ulpicher Stra\ss e 77, 50937 K\"oln, Germany}

   \date{Received 9 January 2026 / Accepted 15 May 2026}

\abstract 
{Stellar mergers  produce 
``explosive outflows" that serve as transient sources of infrared line luminosity and inject mechanical energy early into the natal molecular cloud. 
We present the first velocity-resolved, sub-km\,s$^{-1}$ resolution maps of
the  [\OI]\,63 and 145\,$\upmu$m fine-structure line emission from the wide-angle  outflow in Orion BN/KL, the nearest explosive outflow.
The data were obtained with SOFIA and include new, sensitive  
[\CII]\,158\,$\upmu$m and OH line maps.
They allowed us to disentangle the quiescent cloud gas, traced by a narrow [\OI] component with a full width at half maximum  (FWHM) of \mbox{$\simeq$\,4\,km\,s$^{-1}$}, from the outflow, traced by a broader [\OI] component with a line FWHM of about \mbox{$\simeq$\,20--30\,km\,s$^{-1}$}; the latter exhibits a spatial distribution similar to that of the shock-excited H$_2$ emission seen with JWST.
 The [\OI]\,63\,$\upmu$m line displays
a full width at zero intensity (FWZI) of $\sim$85\,km\,s$^{-1}$ and shows 
foreground narrow absorptions against strong continuum sources. The OH\,119\,$\upmu$m line shows a  prominent \mbox{P-Cygni} profile covering $\sim$160\,km\,s$^{-1}$,
similar to the FWZI of the CO lines.
The total \mbox{[\OI]\,63 and 145\,$\upmu$m} line luminosity  is remarkably high, 
86.5\,\Lsun, of which 55\,\Lsun~is emitted in the broad component.
This luminosity is comparable to the H$_2$ and CO line luminosities,   
 implying an outflow \mbox{mass-loss} rate  of
\mbox{$\dot{M} \simeq (9.1\pm 2.6) \times 10^{-3}$\,\Msun\,yr$^{-1}$}
and a mass \mbox{$M$\,$\simeq$\,(3.3--5.9)\,\Msun}.
 The  \mbox{[\OI]\,63\,/\,145} and \mbox{[\OI]\,63\,/\,[\CII]\,158} intensity ratios reach  very high values in the line wings (20–30 and 40–60, respectively), exceeding those  found in photodissociation regions. These ratios are consistent with the presence of
dense ($n_{\rm H}$\,$\simeq$\,10$^5$ to  10$^6$\,cm$^{-3}$) and warm ($T$\,$\lesssim$\,500\,K) 
\mbox{post-shock} gas. 
We analyzed the fine-structure line-wing intensities using magnetized
shock models that include UV irradiation, to which the \mbox{[\CII] 158\,$\upmu$m}
line intensity is particularly sensitive.
We find that the [\OI] and [\CII] intensities are consistent with emission from dissociative \mbox{$J$-type} shocks with velocities of \mbox{30--40\,km\,s$^{-1}$} and preshock gas densities of a few $10^{4}$\,cm$^{-3}$, illuminated by external UV radiation generated by surrounding fast shocks and possibly by massive 
(proto)stars in the region.
We also report a broad [\OI]\,63\,$\upmu$m  emission feature around the BN star, which we attribute to an unresolved  outflow 
or wind bow shock.}

\maketitle

\section{Introduction}\label{sec:introduction}

Explosive events triggered by mergers of protostars generate transitory infrared
(IR) line \mbox{luminosity} bursts and inject substantial mechanical energy
($\sim$\,10$^{47-48}$\,erg) into the natal molecular cloud, serving as an early source of mechanical feedback in \mbox{clustered} star-forming regions \mbox{\cite[e.g.,][]{Bally05}}.
The resulting \mbox{bow-shock tips}, or ``bullets,''  propagate through the cloud, generating shocks that cool via molecular and atomic line emission, depending on their velocity, magnetic field, gas density, and ambient ultraviolet (UV) radiation.
Stellar mergers must  occur relatively frequently in clustered star-forming regions  \citep[perhaps once per century in our Galaxy; e.g.,][]{Zapata20,Zapata23} and they are therefore likely to be common in luminous galaxies. The Orion BN/KL outflow
\citep[e.g.,][]{Kwan76,Beckwith78,Snell84} is the nearest example of an \mbox{``explosive outflow''}  \citep[e.g.,][]{Gomez05, Zapata09,Goddi11, Bally11, Bally20}. 
Its distinctive kinematics and diverse gas cooling lines
\citep[atomic, H$_2$, CO, H$_2$O, and OH; e.g.,][]{Werner84,Rosenthal00,Gonzalez02,Zapata09,Peng12,Goipacs15}
indicate the presence of shocks, in which the  mechanical energy of the explosion is converted into gas heating and compression as the outflow impacts the swept-up gas and the ambient molecular cloud. However, the nature of these shock(s), `jump'' (\mbox{$J$-type}), ``continuous'' (\mbox{$C$-type}) or mixed, as well as the  role of \mbox{UV radiation}, whether shock-generated or external, remain uncertain. 
On the one hand, \mbox{UV-irradiated} shocks are best traced by velocity-resolved observations of the [\OI] 63 and 145\,$\upmu$m and [\CII]\,158\,$\upmu$m cooling lines, but only a few sources have been mapped in the wings of all three lines. In particular, the scarcity of [\OI]\,145\,$\upmu$m data limits constraints on the physical conditions of the [\OI]-emitting gas across velocities.
On the other hand, only recently have 
shock models \citep[e.g.,][]{Hollenbach79,Hollenbach89,Draine80, Kaufman96} included realistic descriptions of  \mbox{UV-irradiated} shocks \citep[][]{Lesaffre13,Melnick15,Godard19,Godard24,Lehmann20,Lehmann22,Kristensen23}.

Here, we present the highest so far angular resolution [\OI]\,63 and 145\,$\upmu$m maps of the BN/KL outflow (6$''$ and 13$''$, respectively) carried out at sub-km\,s$^{-1}$ resolution and complemented by sensitive line maps of [\CII]\,158\,$\upmu$m, a tracer of UV radiation and stellar feedback \citep[e.g.,][]{Tielens85b,Stacey93,Goicoechea15,Pabst19}, and OH,  a tracer of the outflow kinematics \citep[e.g.,][]{Melnick87,Melnick90,Goico06,Goipacs15}. We obtained these observations with the  
Stratospheric \mbox{Observatory for Infrared Astronomy} \citep[SOFIA;][]{Young12}
using GREAT\footnote{The German REceiver for Astronomy at Terahertz frequencies was developed by the MPI f\"ur Radioastronomie and the KOSMA/Universit\"at zu K\"oln, in cooperation with the DLR Institut f\"ur Optische Sensorsysteme.},
 providing a legacy dataset for studying the  properties of shocked gas.

The paper is organized as follows. 
In   \mbox{Sect.~\ref{sec-intro-orion}}, we provide details on the
 outflow.
In \mbox{Sect.~\ref{sec:observations}}, we describe the airborne observations. In \mbox{Sect.~\ref{sec:results}}, we present the main observational findings, 
 while in \mbox{Sect.~\ref{sec:analysis}}, we detail the physical conditions as a function of velocity in the outflow. Finally, \mbox{Sect.~\ref{sec:discussion}} discusses the nature of the \mbox{[\OI]-emitting} shocks based on comparisons with state-of-the-art shock models and places our results in the context of FIR line emission from other massive star-forming regions.

\section{The Orion BN/KL explosive outflow}\label{sec-intro-orion}

Embedded in the heart of the Orion Molecular Cloud-1 core (\mbox{OMC-1}), just behind the  Orion Nebula (M42), the \mbox{Becklin–Neugebauer/Kleinmann–Low} (BN/KL) region is the closest \citep[$\sim$414\,pc;][]{Menten07} 
 high-mass star-forming region   
\citep[][]{Genzel89,ODell01,Bally08}.
 In addition to the two well-known runaway stars BN and source~$I$, which are the most massive objects in this region  
 \citep[e.g.,][]{Lonsdale82, Scoville83, Menten95,Bally20}, the field hosts the first identified molecular outflow, exhibiting high-velocity CO emission, with line wings extending to over $\pm$100\,km\,s$^{-1}$, together with a  wide-angle, shock-excited H$_2$ outflow
 \citep[e.g.,][]{Kwan76,Beckwith78,Snell84}.
Observations of this outflow have long provided key benchmarks for the development of shock models  \citep[][]{Draine82,Neufeld89b,Hollenbach89}.
 Today we know that this is an \mbox{explosive outflow} produced by a dynamical decay event \mbox{(e.g., a stellar merger)}, which triggered the acceleration of BN and \mbox{source~$I$}, about 500~yr ago \citep[e.g.,][]{Gomez05, Zapata09, Bally11, Goddi11, Bally20}. 
The center of the explosion is located between the current positions of these stars.

\begin{figure}[t]
\centering   
\includegraphics[scale=0.375,angle=0]{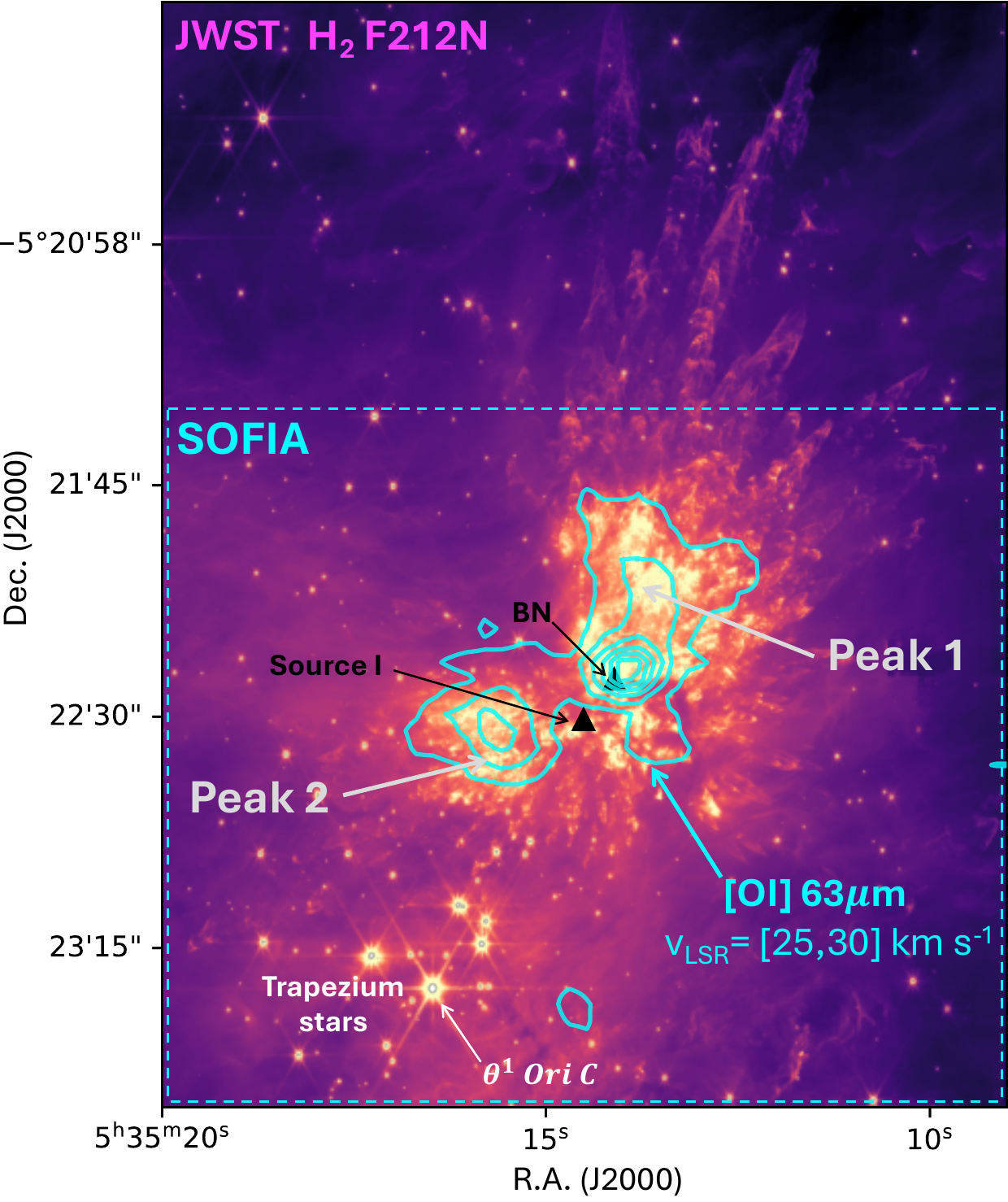}
\caption{BN/KL outflow and the Trapezium cluster observed with JWST/NIRCam H$_2$ F212N \citep{McCaughrean23}. 
 The dashed square marks the field of view mapped with SOFIA/GREAT. Cyan contours show the [\OI]\,63\,$\upmu$m  redshifted line-wing emission over 
 \mbox{$v_{\rm LSR}=25$--30\,km\,s$^{-1}$}, from 15 to 90\,K\,km\,s$^{-1}$ in steps of 15\,K\,km\,s$^{-1}$.}\label{fig:H2} 
\end{figure}

ALMA  CO observations reveal an approximately spherically symmetric \mbox{``Hubble-Lema\^itre flow,''} that is, radial velocities scaling with the projected distance from the  center. The flow is composed of over a hundred narrow \mbox{bow-shock} wakes, or \mbox{``fingers,''}  within $\sim$\,1$'$ ($\sim$0.1\,pc) of the center
\citep{Bally17}. 
Due to the isotropic distribution of the flow, the red- and blueshifted fingers appear to overlap when projected onto the plane of the sky, a characteristic of explosive outflows.
The bullets show shock-excited H$_2$ \mbox{$v$\,=\,1--0} $S$(1) (2.12\,$\upmu$m)  and forbidden [\FeII]\,1.64\,$\upmu$m emission at their tips, together with H$_2$ and lower velocity CO  emission in their wakes
 \citep[e.g.,][]{Zapata09,Kristensen07,Kristensen08,Nissen12,Bally15,Youngblood18}. The H$_2$ emission shows a two-lobe northwest–southeast (NW–SE) orientation \citep[e.g.,][]{Allen93,McCaughrean97}, likely because the other part of the outflow—or its IR emission—is partially obscured by the dense dust ridge along \mbox{OMC-1}, which extends toward the northeast (NE).
The outflow contains  $\sim$\,8\,\Msun\,of molecular gas \citep{Snell84}. About half of this mass is at expansion velocities below 20\,km\,s$^{-1}$. 
The other half belongs to the high-velocity outflow, which exhibits some of the brightest  H$_2$ emissions  in the sky. Peak~1, located $\sim$30$''$ northwest of BN, is the strongest \mbox{H$_2$-emitting} region of the outflow (see \mbox{Fig.~\ref{fig:H2}}).

  The total H$_2$ line luminosity across  the outflow is
   \mbox{$120 \pm 60\,$\Lsun}\,\citep{Rosenthal00}. 
Observations of H$_2$ lines toward Peak~1 reveal excitation temperatures rising from $\sim\,$600\,K in the $v=0$ rotational transitions, to $\sim2500$\,K in the vibrationally excited transitions \citep[][]{Rosenthal00,Youngblood18}. 
Observations suggest a more excited component at $\sim$5000\,K, likely caused by H$_2$ formation pumping after shock dissociation and reformation \citep{Geballe17}.
Peak~1 also exhibits
excited CO and H$_2$O  emission associated with warm, $\sim$\,500\,K, and hot, $\sim$\,2500\,K 
molecular gas \citep[][]{Gonzalez02,Goipacs15}. 
In addition, low angular and spectral resolution FIR observations revealed bright 
[\OI] fine-structure line emission
\citep[e.g.,][]{Werner84,Herrmann97,Lerate06,Goipacs15} while
\mbox{\textit{Herschel}/HIFI} enabled the first velocity-resolved
 [\CII]158\,$\upmu$m maps  of the region
\citep{Goicoechea15,Morris16}.

The BN/KL outflow  also shows intense OH emission, with \mbox{$L_{\rm OH}/L_{\rm H_2O} \simeq 0.4$}, 
suggesting ongoing H$_2$O photodissociation \citep[][]{Goipacs15}.
 The low-lying OH rotational lines   fall at FIR wavelengths, where the dust  continuum  is very intense.
These lines require very high  densities to be collisionally excited
(\mbox{$n_{\rm cr}$\,$\gtrsim$\,10$^{8-9}$\,cm$^{-3}$)} and are typically optically thick.
As gas densities are typically lower, their excitation temperatures ($T_{\rm ex}$) often fall below the continuum temperature, producing 
``\mbox{P-Cygni}'' profiles in molecular outflows \citep[first detected toward BN/KL;][]{Betz89,Melnick90,Goico06}.

\section{SOFIA airborne observations  and data reduction}\label{sec:observations}

We mapped multiple FIR lines with GREAT  
 during eight SOFIA flights between January 2014 and February 2021. Apart from the flights taking place in in spring 2021 (observing cycle~8), which we operated out of Cologne-Bonn (CGN, Germany) during the pandemic, we carried out all flights from Palmdale, California. We conducted all observations during GREAT Consortium time (project IDs: 83$\_$0004 (cycle~1), 83$\_$0428 (cycle~4), and 83$\_$0630 \mbox{(cycles~8–9)}. The flights accumulated 5 hours of observing time. 
 
In 2014, we carried out the first observations of the \32OH line (2514.3\,GHz\,$\simeq$\,119.2\,$\upmu$m) with the \mbox{single-pixel} detector of GREAT \citep{Heyminck2012}. In 2021, we extended the map using the corresponding channel of the multi-color 4GREAT receiver \citep{Duran2021}. We observed the [\OI] \mbox{$^3$P$_1$--$^3$P$_2$} line  (4744.8\,GHz\,$\simeq$\,63\,$\upmu$m) with the \mbox{7-pixel} high-frequency array (HFA) of upGREAT \citep{Risacher2018}. We simultaneously observed the [\CII] \mbox{$^2$P$_{3/2}$--$^2$P$_{1/2}$} line  (1900.5\,GHz\,$\simeq$\,158\,$\upmu$m) and the  [\OI] \mbox{$^3$P$_0$--$^3$P$_1$} line (2060.1\,GHz\,$\simeq$\,145\,$\upmu$m) with the two polarization-split upGREAT low-frequency sub-arrays (LFA): the \mbox{7-pixel} \mbox{V-polarization} array tuned to the [\OI] line and the H-polarization array to the [\CII] line. This configuration processed 21 signals in parallel. We employed Fast Fourier Transform spectrometers, updated from \citet{Klein2012}, which provided 32\,k channels across the                        $\sim$0.5–4\,GHz intermediate-frequency bands.
\mbox{Table~\ref{tab:spectral-lines}}  summarizes the instrument configurations.

The instrument and observatory performance (flight altitude, precipitable water vapor)  determined the actual observing mode and source coverage (map sizes), while previous 
\mbox{\textit{Herschel}} spectroscopic maps of \citet{Goipacs15} guided them.
We reference all observations to the nominal source position 
\mbox{RA\,=\,05$^{\rm h}$35$^{\rm m}$14.3$^{\rm s}$}, 
\mbox{Dec\,=\,$-$05$^\circ$22$'$33.7$''$} (J2000) 
in the hot core region. We performed the upGREAT observations of both [\OI] lines and [\CII] in fast total power on-the-fly slewing mode (0.4\,sec integration time per dump), using 3$''$ sampling in both RA and Dec We observed the source twice, scanning once in Dec and once in RA, with the fine sampling driven by the 6.3$''$ beam of the HFA. The map size \mbox{(144$''$$\times$126$''$)} refers to the on-sky coverage of the central array pixels, while the hexagonal array’s outer pixels provide a larger coverage. We selected a ``clean'' on-sky reference position at \mbox{$\Delta$RA, $\Delta$Dec}=\mbox{($-$1700$''$, +900$''$)}. To achieve the most uniform sampling, we tilted the array axis by 19.1$^{\circ}$ relative to the scanning direction \citep[][]{Risacher2016b}.
We observed the OH~$^{2}\Pi_{1/2}$ (1834.7\,GHz\,$\simeq$\,163.4\,$\upmu$m) line during the LFA commissioning in 2016, using 
chopped (1\,Hz) on-the-fly mode with 6$''$ spatial sampling. 
The reference OH  line frequencies (Table~\ref{tab:spectral-lines}) correspond to the line-strength-weighted frequency of each hyperfine-structure triplet, which are spectrally unresolved.
 We observed the OH\,163.4\,$\upmu$m line of  the \mbox{$\Lambda$-doublet} in the 
upper sideband,
thereby avoiding to blend with the 163.1\,$\upmu$m line of the doublet in our
dual sideband  receiver.

We applied a large chop throw of 240$''$ toward negative RA offsets. Observing the OH~$^{2}\Pi_{3/2}$\,119.2\,$\upmu$m line proved most challenging. At this frequency, the shortage of local oscillator power needed to pump the HEB mixers forced us to couple the signal via a Martin–Puplett interferometer \citep{Heyminck2012}; this, in turn, limited the receiving bandwidth of the sky signal to $\sim$1.6\,GHz.
Thus, the expected velocity coverage of the outflow barely fits into the effective bandpass ($\sim$190\,km\,s$^{-1}$). 
This required pointed (raster) observations, in double-beam chopped mode, for optimal system stability (wobbler throw 300$''$, at $-$20$^{\circ}$ counter-clockwise against positive RA). In 2014, an exploratory 5$\times$5 raster on a 10$''$ grid was observed (with the single-pixel M-channel of GREAT), showing the emission to be compact; in 2021 with 4GREAT we added a \mbox{5$\times$5} raster on a 6$''$ grid. 

The GREAT consortium derived the half-power main beam sizes and efficiencies from planet observations. 
Telescope operators established the pointing on nearby optical reference stars with an accuracy of 1–2$''$. At the beginning of each flight series, we aligned the instrument’s optical axis with these imagers by observing planets. Although the co-alignment between the central pixels is better than 1–2$''$ (and the data header accounts for it), we generally tracked the highest frequency channel in a given configuration—typically the central HFA pixel in most flights.
We amplitude-calibrated the raw data with the KOSMA \mbox{{\it kalibrate}} software \citep{Guan2012}. We corrected the data for atmospheric extinction and calibrated them in 
$T_{\rm mb}$\footnote{We shall refer to the \mbox{``line luminosity''} ($L$)
in erg\,s$^{-1}$ or \Lsun~units. The conversion from integrated  
line intensity (\mbox{$W$\,=\,$\int \Delta T_{\rm mb}\, dv$}) in \mbox{K\,km\,s$^{-1}$}, where $\Delta T_{\rm mb}$ is the 
continuum-subtracted main brightness temperature,
to integrated line intensity ($I$)  
in \mbox{erg\,s$^{-1}$\,cm$^{-2}$\,sr$^{-1}$}
is \mbox{$I$\,=\,2$k$\,$W$\,$\nu^3/c^3$}, or
\mbox{$I$(erg\,s$^{-1}$\,cm$^{-2}$\,sr$^{-1}$)\,=\,$W$(K\,km\,s$^{-1}$)$\cdot$$\nu$(GHz)$^3$\,/\,9.76$\cdot$10$^{14}$}.}. 
When SOFIA flew between 
\mbox{12.2--13.4\,km} in altitude, the atmospheric transmission remained smooth near the target velocities. 

\begin{figure*}[ht]
\centering   
\hspace{0.6cm} 
\includegraphics[scale=0.53,angle=0]{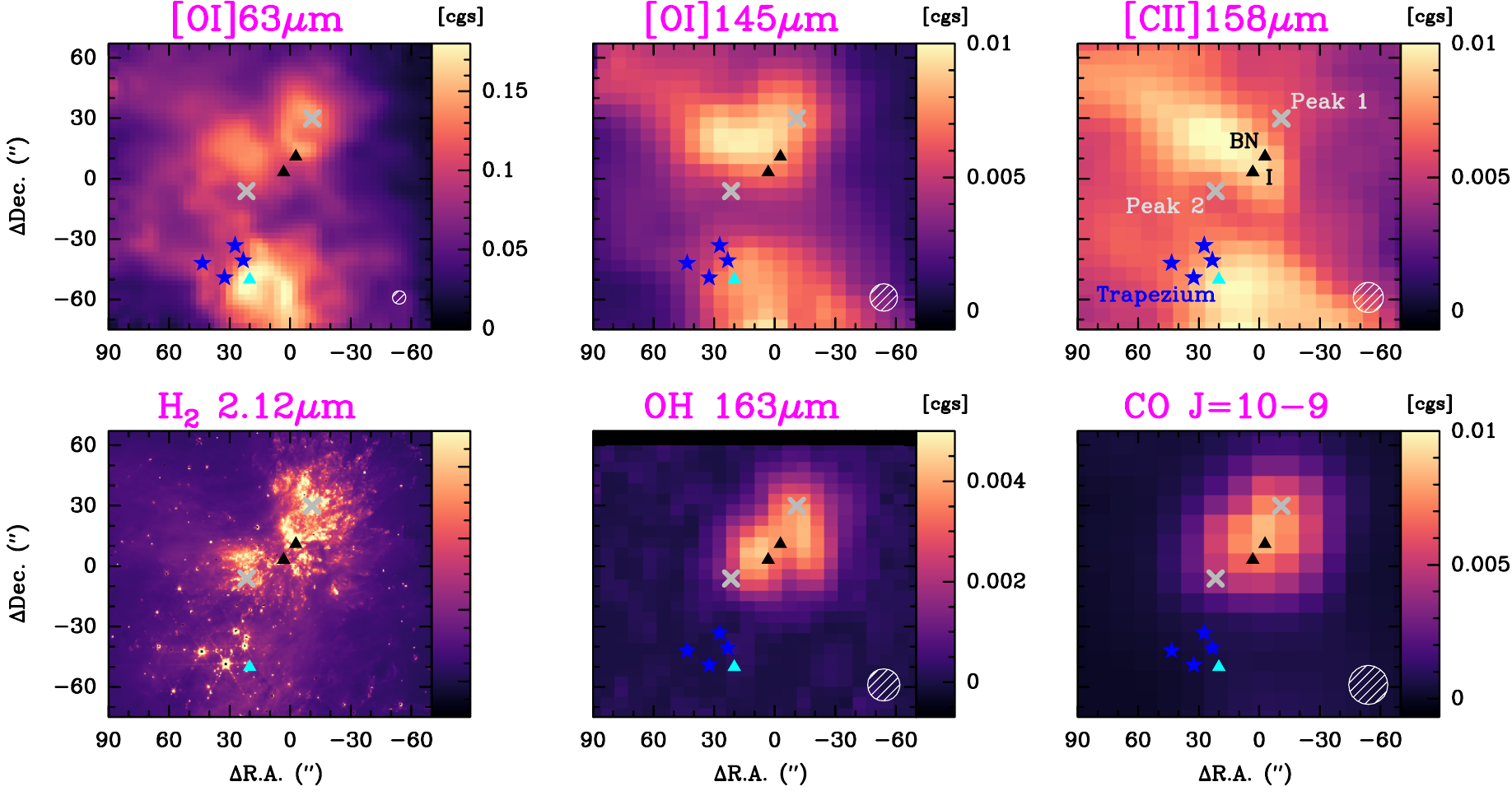} 
\caption{Total line intensity maps 
(in \mbox{erg\,s$^{-1}$\,cm$^{-2}$\,sr$^{-1}$\,=\,``cgs''})  integrated over the
complete line profile.  The beam size  is indicated in the bottom-right corner of each panel except for H$_2$, where it is too small to display (0.1$''$).
 The cyan triangle shows a position near the Trapezium, where the
[\OI]\,63$\upmu$m line intensity peaks (Table~\ref{Table_int_Trapezium}).
The H$_2$ image refers to the JWST/NIRCam F212N image 
\citep[][]{McCaughrean23}.
The CO~$J$\,=\,10-9 map was obtained with \textit{Herschel}/HIFI at 20$''$ resolution \citep[][]{Goico19}.}
\label{fig:sofia_maps_int}
\end{figure*}

\begin{figure*}[t]
\centering  
\hspace{-0.0cm} 
\includegraphics[scale=0.73,angle=0]{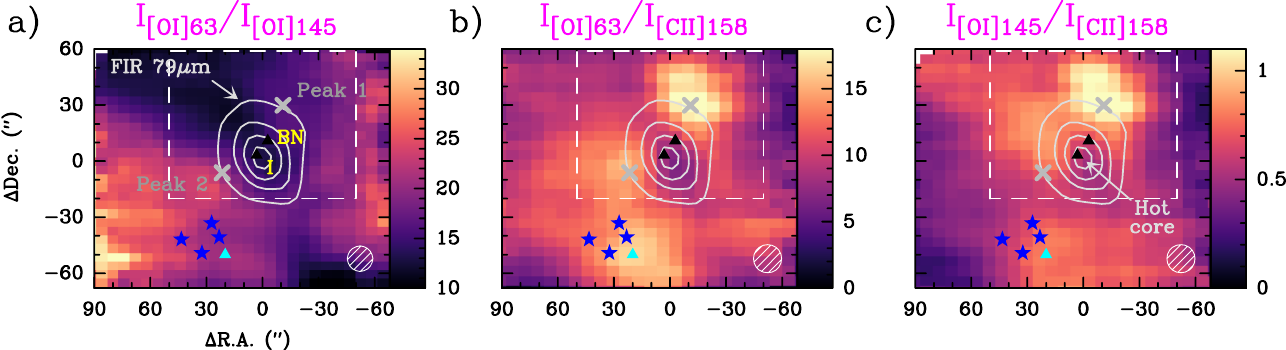}
\caption{Total line intensity ratio maps (derived from line intensities  in \mbox{erg\,s$^{-1}$\,cm$^{-2}$\,sr$^{-1}$}). (a)~\mbox{[\OI]\,63/145} at a common angular resolution of 13$''$, 
(b)~\mbox{[\OI]\,63/[\CII]\,158} at 15$''$, and (c)
\mbox{[\OI]\,145/\,[\CII]158} at 15$''$. The dashed box shows the \mbox{(100$''\times$80$''$)} area used to extract both the line luminosities in the outflow region 
\mbox{(Table~\ref{Table_intensities})} and the line intensities as a function of velocity 
(see Figs.~\ref{fig:wing_ratios} and \ref{fig:model_MTC_outflow}).  
Contours show the FIR 79\,$\upmu$m-continuum obtained with \textit{Herschel}/PACS, from 2.5 to 10 (10$^3$ Jy), centered at the position of the hot core \citep[][]{Goipacs15}.}
\label{fig:sofia_maps_int_ratios}
\end{figure*}

We  processed the calibrated data with  GILDAS\footnote{see http://www.iram.fr/IRAMFR/GILDAS.}. We  inspected the data to remove noisy spectra and other artifacts (e.g., spikes from onboard interference). 
We smoothed the spectra to a channel resolution of 0.15\,km\,s$^{-1}$ (0.3\,km\,s$^{-1}$ for the  OH~119\,$\upmu$m line) and removed first-order baselines.
The spectra were then gridded into a data cube through a convolution with a Gaussian kernel of a full width at half maximum (FWHM) of $\sim$\,1/3 of the telescope beamwidth at the different frequencies.
 The 1$\sigma$ rms noise  in the maps are $\sim$3.5\,K ([\OI]\,63\,$\upmu$m), $\sim$2.5\,K ([\OI]\,145\,$\upmu$m),
and $\sim$2.0\,K ([\CII]\,158\,$\upmu$m), and $\sim$4.3\,K  \mbox{(OH\,163\,$\upmu$m)} per  velocity channel
(as provided by the NOISE routine).
To determine the  noise over a given velocity bin, we used
\mbox{$\sigma_{\rm bin} = \sigma_{\rm ch} \, \delta v_{\rm ch} \, \sqrt{N_{\rm ch}}$},
where $\sigma_{\rm ch}$ is the rms per channel, $\delta v_{\rm ch}$ the channel width,
and $N_{\rm ch}$ the number of channels in that bin. For specific applications (e.g., line intensity ratio maps), 
we convolved the data cubes with Gaussian kernels to a common angular resolution, corresponding to FWHMs of 13$''$ or 15$''$.

\section{Results}\label{sec:results}

Figure~\ref{fig:H2} shows the core of \mbox{OMC-1} observed with  JWST/NIRCam  F212N \citep[][]{McCaughrean23}, dominated by H$_2$ \mbox{$v$\,=\,1--0} $S$(1) line
emission at 2.12\,$\upmu$m.  
The dashed square marks the field of view mapped with SOFIA/GREAT, encompassing the wide-angle BN/KL outflow, including the bright Peak~1 and Peak~2 regions.
The cyan contours in \mbox{Fig.~\ref{fig:H2}} show the [\OI]\,63\,$\upmu$m redshifted line-wing emission at $\sim$20\,km\,s$^{-1}$ relative to the cloud systemic velocity of \mbox{$\simeq$8--9\,km\,s$^{-1}$} \citep[hereafter \mbox{$v_{\rm LSR,\,0}$;}][]{Bally87,Berne14}. This [\OI] emission peaks
 close to BN (\mbox{Sect.~\ref{sec:compact-outflow}})
and follows the H$_2$ emission from the outflow.
The field also includes the massive stars of the Trapezium cluster, in the foreground, located at 
$\sim$0.2\,pc in front of the cloud. Their far-UV (FUV; \mbox{$6 < E < 13.6$\,eV}) 
radiation, dominated by $\theta^1$~Ori~C star, illuminates all surfaces  of \mbox{OMC-1}.

\begin{figure*}[ht]
\centering   
\includegraphics[scale=0.374,angle=0]{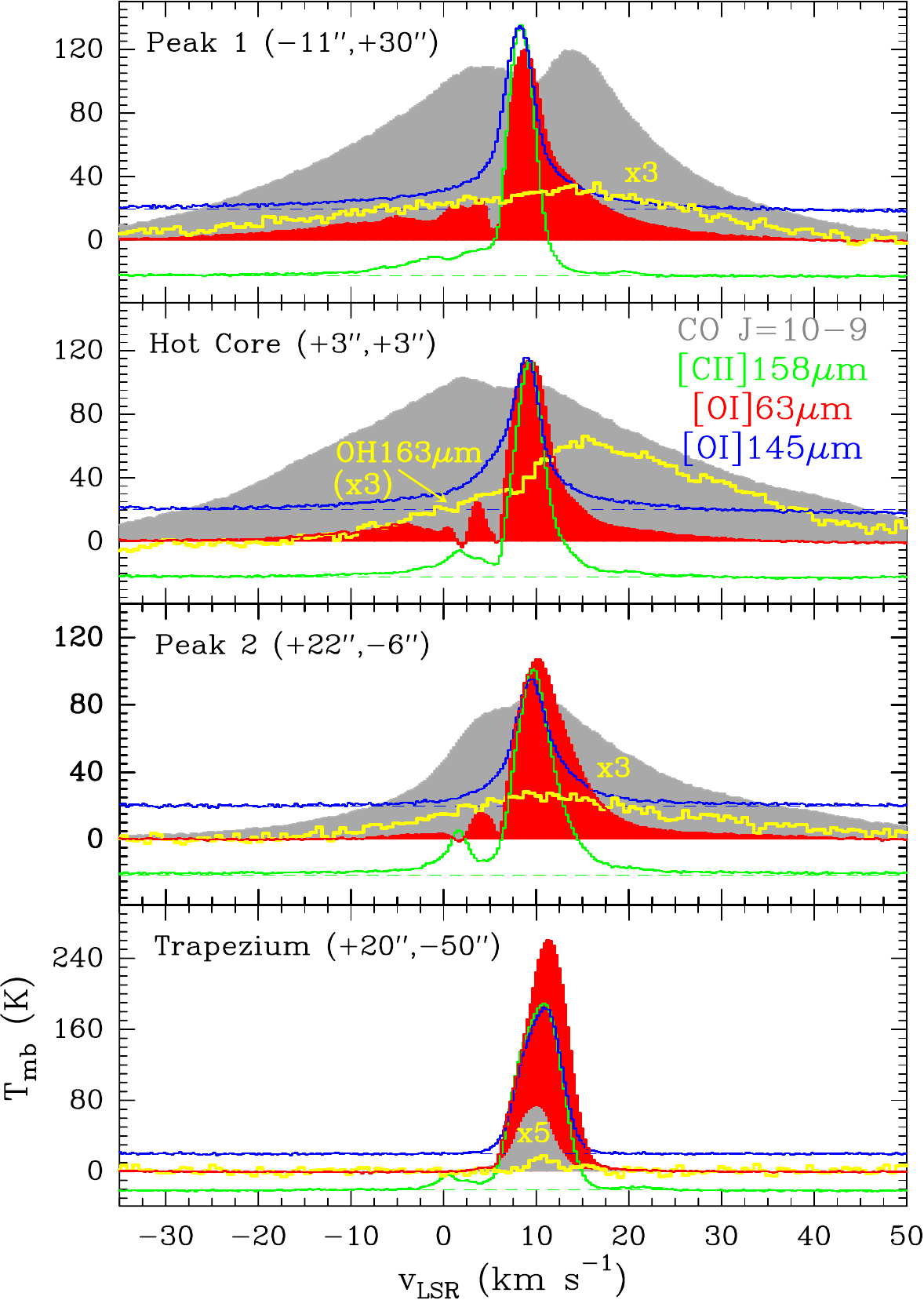} 
\hspace{2cm}
\includegraphics[scale=0.374,angle=0]{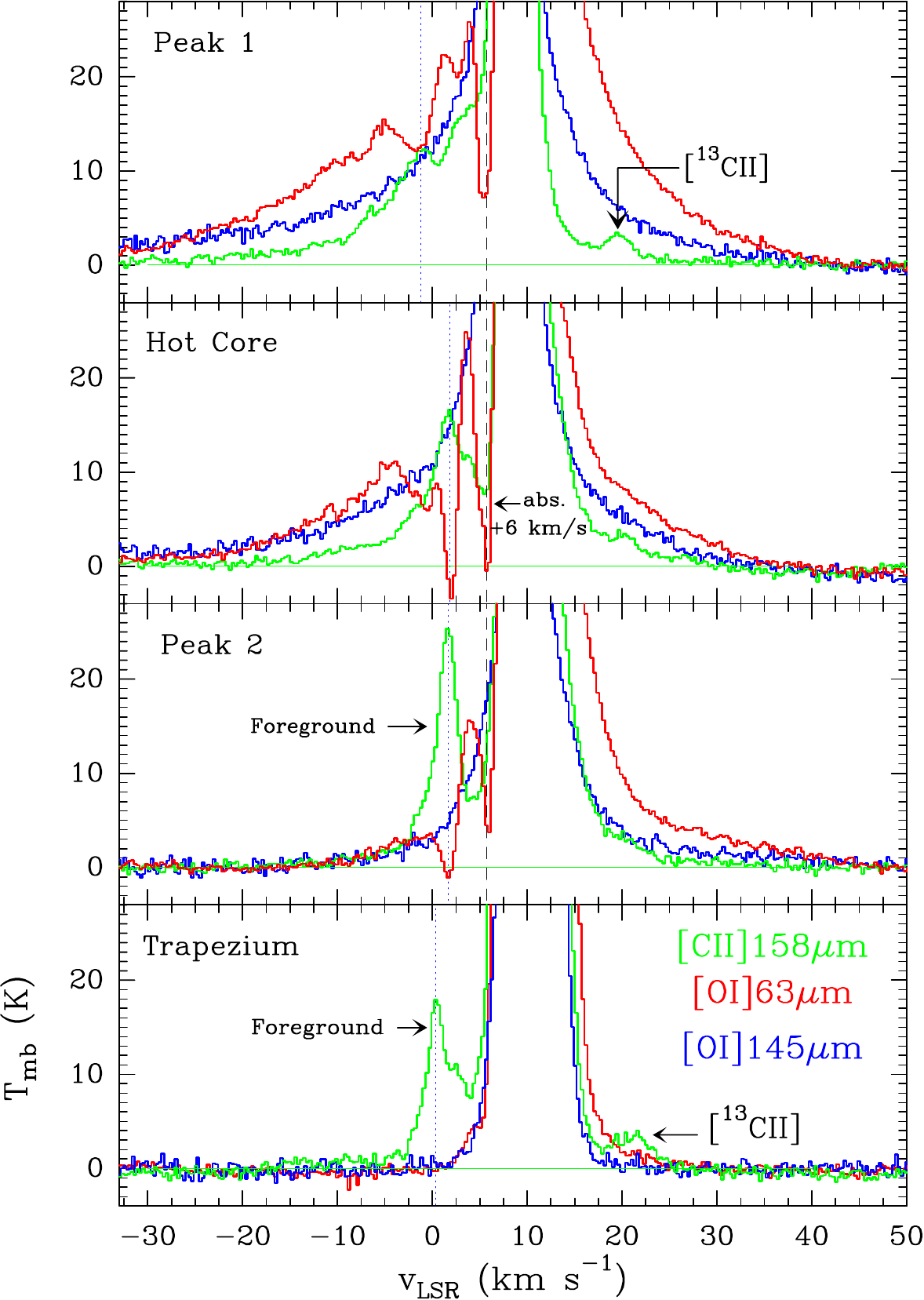} 
\caption{
Velocity-resolved spectra at representative positions, all from maps convolved to 15$''$ (except CO 10--9, with a beam of 20$''$). 
\textit{Left:} Complete spectra, with offsets in arcseconds given in parenthesis; continuum levels of [\CII] 158\,$\upmu$m and [\OI] 145\,$\upmu$m shifted for clarity. 
The OH\,163\,$\upmu$m emission lines (yellow) are scaled by a factor of three
or five. 
\textit{Right:} Zoom on faint line wings and foreground line features.}
\label{fig:spectra}
\end{figure*}

\subsection{Integrated line  intensity maps}

Figure~\ref{fig:sofia_maps_int} shows [\OI]\,63,145\,$\upmu$m, [\CII]\,158\,$\upmu$m,
and OH\,163\,$\upmu$m total line intensity maps (integrated over their complete line profile) in units of \mbox{erg\,s$^{-1}$\,cm$^{-2}$\,sr$^{-1}$}, which are used to compare the emitted line luminosity and line cooling budget. In these maps we mark the positions of key stars (BN and
source~$I$) and environments (Peak~1, Peak~2, and the Trapezium region), which we study throughout the manuscript.
The [\OI]\,63\,$\upmu$m emission is the brightest FIR atomic fine-structure line in the region and roughly follows the H$_2$ emission along the wide-angle outflow. 
Still, the line intensity peaks southwest of the Trapezium stars, hereafter  the ``Trapezium position,'' at \mbox{($+$20$''$, $-$50$''$)} from the map center (see details in \mbox{Sect.~\ref{sub-sec:profiles}}). 
This region is part of the large-scale, face-on photodissociation region
(PDR), with $n_{\rm H}$\,$\gtrsim$\,10$^{5}$\,cm$^{-3}$ 
\mbox{\citep[e.g.,][]{Herrmann97,Goicoechea15,Goico19}},
at the interfaces between \mbox{OMC-1} 
 and the foreground  \HII~region, which is photoionized by 
\mbox{$\theta^1$\,Ori\,C}. This PDR component
is located closer to the observer than the BN/KL outflow and
 produces  bright narrow  line emission---often called the ``spike'' component---at nearly all positions in the field.

The [\OI]\,145\,$\upmu$m line is strikingly bright and as bright as the [\CII]\,158\,$\upmu$m line, displaying a similar spatial distribution as the [\OI]\,63\,$\upmu$m emission.
As observed before  \mbox{\citep[e.g.,][]{Goicoechea15,Morris16}}, the [\CII]\,158\,$\upmu$m integrated intensity map does not follow the H$_2$ emission. Instead, [\CII]\,158\,$\upmu$m  shows strong emission close to the Trapezium, along with lower line intensity  emission  from the
FUV-irradiated surface of \mbox{OMC-1}.
The \mbox{\12OH} line emission at  163 $\upmu$m is confined to the outflow, peaking toward the inner hot core region, with only very faint emission from  the quiescent cloud
and PDR. 
The spatial distribution of the OH\,163\,$\upmu$m emission resembles that of the 
 high-$J$ CO emission 
\citep[e.g.,][]{Peng12,Goipacs15,Goico19}. 

Figure~\ref{fig:sofia_maps_int_ratios} shows  the atomic fine-structure  line intensity ratio maps (integrated over the entire line profile). 
Typically, these intensity ratios provide diagnostics of the gas density, temperature, and FUV radiation field strength in the warm gas associated with PDRs and shocks.
Although these  maps do not reveal the velocity structure of the  gas, they display a different spatial distribution compared to the absolute line intensity maps (Fig.~\ref{fig:sofia_maps_int}). 
 The \mbox{[\OI]\,63/[\CII]\,158} intensity ratio map shows a clear maximum ($\simeq$\,20) toward  Peak~1  (\mbox{Fig.~\ref{fig:sofia_maps_int_ratios}b}). The \mbox{[\OI]\,145/[\CII]\,158}  map exhibits a similar spatial distribution, with a high ratio ($\gtrsim$1) toward Peak~1 (\mbox{Fig.~\ref{fig:sofia_maps_int_ratios}c}). 
Given the much higher critical densities for collisional excitation of the [\OI]\,63 and 145\,$\upmu$m lines compared to that of the 
[\CII]\,158\,$\upmu$m line (\mbox{see Table~\ref{tab:spectral-lines}}), these peaks trace the highest density gas (assuming similar FUV illumination).
In general, the \mbox{[\OI]\,63\,/\,[\CII]\,158\,$\upmu$m} line intensity ratios toward 
Peak~1 are higher than those observed toward protostellar outflows
\citep[e.g.,][]{Liseau06}  and PDRs 
\mbox{\citep[e.g.,][]{Salas12}}.
\mbox{Figure~\ref{fig:sofia_maps_int_ratios}} (contours) also shows the FIR continuum map at 79\,$\upmu$m  \citep{Goicoechea15}, tracing the extended warm dust emission from the hot core region. The continuum emission is fainter toward the Peak~1 and Peak~2.
The \mbox{[\OI]\,63\,/\,145} ratio shows its lowest values ($\simeq$10) in a region roughly orthogonal to the wide-angle outflow (\mbox{Fig.~\ref{fig:sofia_maps_int_ratios}a}), following the quiescent cloud, the ``extended ridge.'' 
Radiative transfer models (\mbox{Sect.~\ref{sub-sec:MTC-models}}) indicate that the low intensity ratios are consistent with the [\OI]\,63\,$\upmu$m line being optically thick, and with the gas having lower density than in the post-shock gas.

\subsection{Velocity-resolved far-IR line profiles} \label{sub-sec:profiles}

\subsubsection{Line wings}

The GREAT maps enable us to resolve the velocity structure of the 
[\OI]\,63 and 145\,$\upmu$m emission  and to compare it with other
FIR lines: [\CII]\,158\,$\upmu$m, OH, and high-$J$ CO. \mbox{Figure~\ref{fig:spectra}} presents such a comparison for four representative positions: Peak~1, the hot core region, Peak~2, and the Trapezium position.
The first three positions lie on the outflow region and show  line wing 
emission.
As is apparent from these spectra, the OH and CO lines exhibit 
similarly broad line profiles, with $\Delta v_{\rm FWHM}$\,$\gtrsim$\,30\,km\,s$^{-1}$.
These lines reveal prominent wings, with full widths at zero intensity (FWZI) exceeding 150\,km\,s$^{-1}$ \citep[see \mbox{Fig.~\ref{fig:Figure_oh119}} and][for additional velocity-resolved CO studies]{Kwan76,Zapata09,Peng12}.
 These line profiles differ markedly from those of the FIR atomic fine-structure lines
 (\mbox{Fig.~\ref{fig:spectra}}), which consist  of a bright narrow component, \mbox{$\Delta v_{\rm FWHM} \simeq$\,4\,km\,s$^{-1}$}, together with moderately
broad wings. This suggests that the shocks producing the atomic fine-structure emission differ in nature or location from those driving the highest velocity CO and OH emission.
 
The FWZI of the \mbox{[\OI]\,63} and 145\,$\upmu$m lines toward Peak~1 is $\sim$85\,km\,s$^{-1}$ and $\sim$62\,km\,s$^{-1}$, respectively.
The [\OI] FWZIs exceed that of the [\CII]\,158\,$\upmu$m line \mbox{($\lesssim$\,50\,km\,s$^{-1}$)}. This implies that the highest velocity component either lacks FUV illumination or that C$^+$ is rapidly converted into other species 
in the post-shock gas \mbox{\citep[see also][]{Morris16}}. 
The [\OI]\,145\,$\upmu$m line is optically thin and is not affected
by foreground absorptions (see \mbox{Sect.~\ref{subsec-foreground}}). Toward Peak~1, the [\OI]\,145\,$\upmu$m line profile  can 
 be  fitted with two Gaussians of similar intensity but different line FWHM:
a narrow one (\mbox{$\Delta v_{\rm FWHM} \simeq 3.6 \pm 0.1$\,km\,s$^{-1}$} 
 centered at \mbox{$v_{\rm LSR} \simeq +8.2 \pm 0.1$\,km\,s$^{-1}$}),  consistent with gas in the quiscent face-on PDR, and a broad one (\mbox{$\Delta v_{\rm FWHM} \simeq 19.6 \pm 0.3$\,km\,s$^{-1}$} at \mbox{$v_{\rm LSR} \simeq +7.4 \pm 0.1$\,km\,s$^{-1}$}), implying supersonic velocity
dispersions associated with shocked gas.

In Appendix~\ref{App:channel_maps} we show the velocity channel maps of the observed FIR lines. As in \mbox{Fig.~\ref{fig:H2}}, the redshifted [\OI]\,63 and 145\,$\upmu$m emission follows the H$_2$ and CO emission along the two lobes of the wide-angle outflow, whereas the redshifted [\CII]\,158\,$\upmu$m emission is weak toward Peak~1 and Peak~2 (Fig.~\ref{fig:channels}). The [\OI]\,63 and 145\,$\upmu$m emission from the blueshifted wing peaks  around BN and  Peak~1.  At $v_{\rm LSR,0}$, the atomic lines  show extended emission in the SW–NE direction, orthogonal to the wide-angle outflow, mostly tracing 
the illuminated rim of \mbox{OMC-1}'s ridge.

Assuming that most of the narrow-line emission arises from the face-on PDR at the rims of \mbox{OMC-1}, as suggested by 
its spatial distribution, we estimate that  $\sim$65\%, $\sim$50\%, and $\sim$15\% of the observed 
[\OI]\,63, [\OI]\,145, and [\CII]\,158\,$\upmu$m emissions, respectively, originate from  shocked gas. Nonetheless, some of the narrow line emission may originate from low-velocity outflows or from shocks oblique to the line of sight, so the above percentages should be considered lower limits. 

\subsubsection{[\OI]\,63\,$\upmu$m narrow-line ``spike'' emission}

For bright, optically thick [\OI]\,63\,$\upmu$m emission, the continuum-subtracted peak main-beam temperature, \mbox{$T_{\rm P}=J(T_{\rm ex})-J(T_{\rm dust})$}, where $J(T)$ is the Planck-corrected radiation temperature, yields the excitation temperature, $T_{\rm ex}$, of the line, arising here  from the extended PDR in front of the outflow:
 \begin{equation} 
T_{\rm ex}({\rm [OI]}\,63\,\upmu\mathrm{m}) = \frac{227.7}{\ln\!\left(1 + \frac{227.7}{T_{\rm P}({\rm [OI]}\,63\,\upmu\mathrm{m}) + J(T_{\rm dust})}\right)}\ \mathrm{K},
\end{equation}\label{eq-Tex}where \mbox{$\Delta E$/$k_{\rm B}$\,=\,$(E_{\rm u} - E_{\rm l})$/$k_{\rm B}$\,=\, 227.7\,K}, and 
\mbox{$J(T_{\rm dust})$\,=\,$T_{\rm c}$} is the continuum brightness temperature at 63\,$\upmu$m.
In most observed positions, 
\mbox{$T_{\rm P}$([OI]\,63\,$\upmu$m)\,$\simeq$\,100--120\,K}, and 
\mbox{$J(T_{\rm dust})$\,$\simeq$\,5--100\,K}
\citep[from the 63\,$\upmu$m continuum levels of][]{Werner84}, where the
highest value refers to the hot core region. These
\mbox{$T_{\rm P}$([OI]\,63\,$\upmu$m)}
values translate to
\mbox{$T_{\rm ex}$\,$\gtrsim$\,200\,K}, which is a lower
limit to the gas temperature (e.g., \mbox{Fig.~\ref{fig:oi_Tex}}).

Far from the outflow, the atomic fine-structure lines are \mbox{narrow} and lack high-velocity wings.
This is exemplified by the spectra toward the Trapezium position (\mbox{Fig.~\ref{fig:spectra}}), where these lines show Gaussian profiles and line
FWHM of about 5\,km\,s$^{-1}$ \mbox{(Table~\ref{Table_int_Trapezium})}. 
Still, this particular position—resembling a cavity structure—shows
remarkably strong \mbox{[\OI]\,63\,$\upmu$m} emission, with a peak brightness
temperature of $T_{\rm peak}=330$\,K (at the native 6$''$ resolution).
This implies $T_{\rm k} \geq T_{\rm ex} \simeq 440$\,K (with $T_{\rm c}$\,$\simeq$\,5\,K, and assuming a beam filling of one), consistent with warm-to-hot
gas in a dense PDR facing the Trapezium. 
This excited PDR is located near the region of maximum \mbox{CH$^+$ $J$\,=\,1--0} emission in \mbox{OMC-1} \citep[\mbox{position~\#3} of][]{Goico19}, a signature of strong FUV irradiation and elevated temperatures.
JWST images reveal this and similar  cavities illuminated by FUV radiation from the Trapezium; e.g., \mbox{Fig.~4} of \citet{McCaughrean23} and \mbox{Fig.~4} of \citet{Habart24}.
These \mbox{``crenellations''} \citep{ODell15} at \mbox{OMC-1}'s illuminated surface were likely carved by protostellar outflows \citep[see also][]{Kavak22}.

 \subsubsection{Foreground [\OI]\,63\,$\upmu$m line absorption}
 \label{subsec-foreground}

The [\OI]\,63\,$\upmu$m line is a ground-state transition.
The strong  dust continuum at 63\,$\upmu$m from the central BN/KL region
\citep[][]{Werner76,Cerni06,Goicoechea15}, together with the presence of
low excitation gas along the line of sight, produces absorption features
in the blueshifted [\OI]\,63\,$\upmu$m line wing, where \mbox{$T_{\rm ex}<T_{\rm c}$}.
 In this region, the [\OI]\,63\,$\upmu$m line profile shows a narrow, a few km\,s$^{-1}$
 FWHM, absorption component at \mbox{$v_{\rm LSR} \simeq +6$\,km\,s$^{-1}$}, which is typically associated with gas in the hot core itself 
\citep[e.g.,][]{Blake87,Tercero10}. In addition, it shows blueshifted absorptions, at \mbox{$v_{\rm LSR} \simeq -1$\,km\,s$^{-1}$} (toward Peak~1) and at \mbox{$v_{\rm LSR} \simeq +2$\,km\,s$^{-1}$} (toward the hot core and Peak~2).
These blueshifted absorptions coincide with [\CII]\,158\,$\upmu$m emission peaks 
(left panels of \mbox{Fig.~\ref{fig:spectra}}) 
known to arise from foreground material in Orion's Veil 
\citep[][]{Goicoechea15,Pabst19}, including the \mbox{``Northern Dark Lane”} and the 
\mbox{``Dark Bay,”} which are apparent in optical images \citep[][]{ODell01}.
\mbox{Because} of the lower density in these   components (\mbox{$n_{\rm H} \approx 10^3$\,cm$^{-3}$}; \citealt{Pabst20}), the fainter 158\,$\upmu$m continuum, and the much lower critical density  of the [\CII]\,158\,$\upmu$m transition, leading to \mbox{$T_{\rm ex} > T_{\rm c}$}, these  foreground  gas layers appear as [\CII]\,158\,$\upmu$m emission components. 
Away from the hot core region, the [\OI] 63\,$\upmu$m line profiles show no
absorption components.

\begin{figure}[t]
\centering   
\includegraphics[scale=0.33,angle=0]{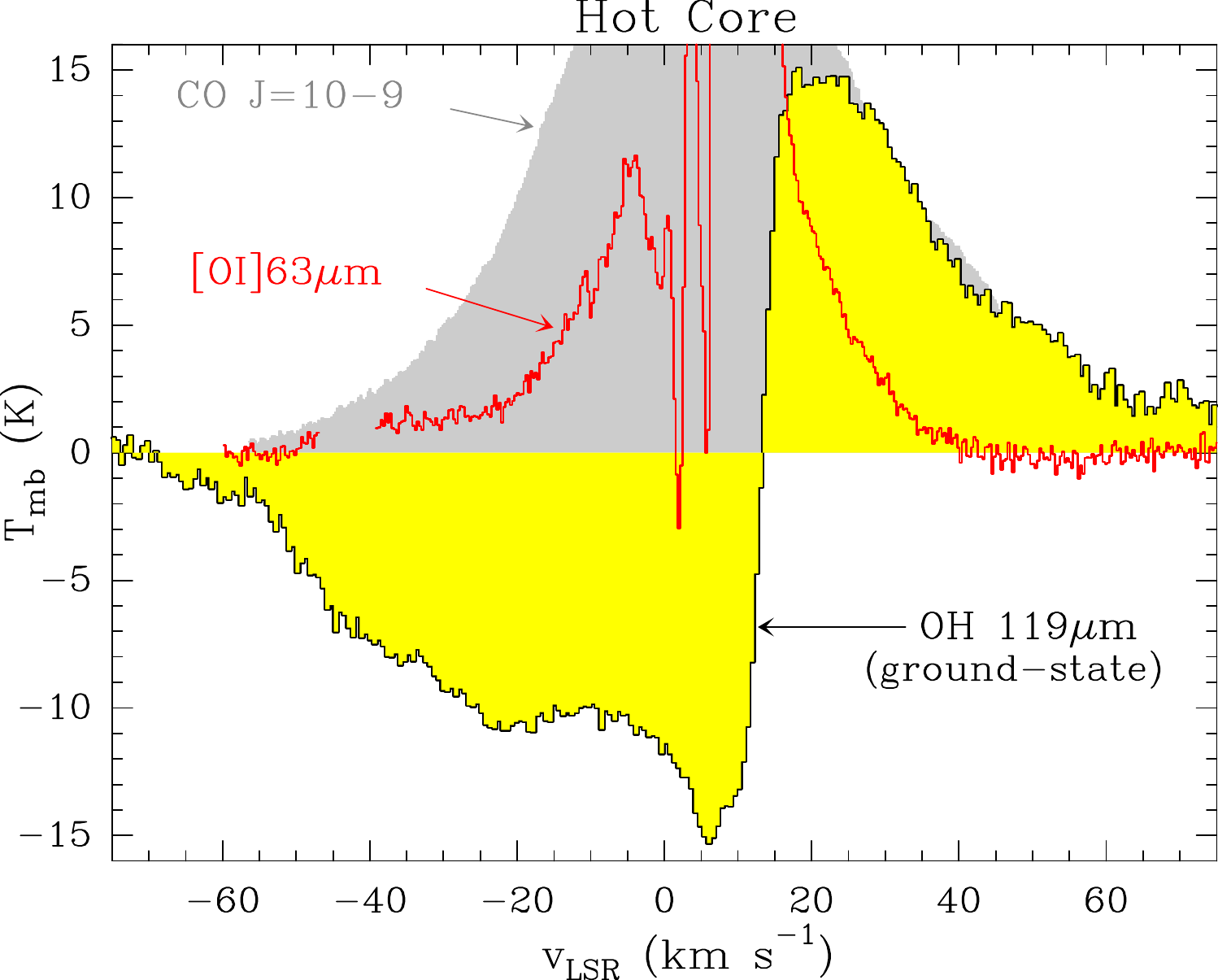}
\caption{P-Cygni profile of the OH\,119\,$\upmu$m line toward the hot core region (the FIR continuum peak), compared to other line profiles. See Fig.~\ref{fig:ZOOM_ALL_OH} for a zoom on the low-intensity features.}  
\label{fig:Figure_oh119}
\end{figure}

\begin{figure*}[ht] 
\centering   
\includegraphics[scale=0.46,angle=0]{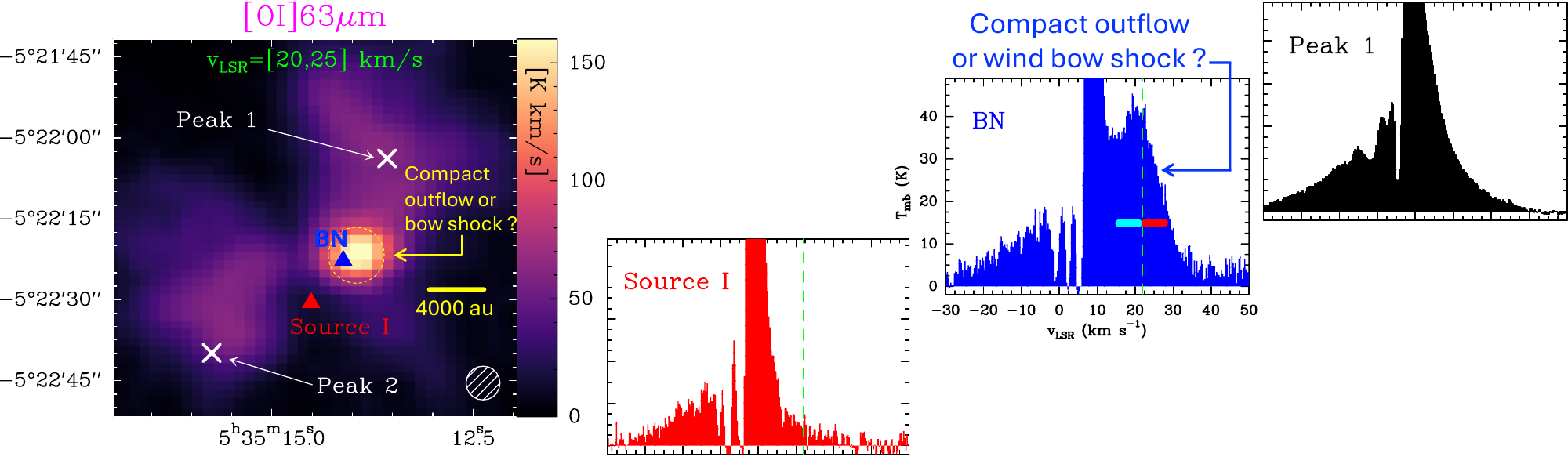}
\caption{Possible compact outflow or wind bow shock around BN star. \textit{Left}: Spatial distribution of the  redshifted [\OI]\,63\,$\upmu$m emission
in the $v_{\rm LSR}$ range from $+$20 to $+$25\,km\,s$^{-1}$. This map (in R.A. and Dec.) shows
 extended emission following the wide-angle H$_2$ outflow  and
a possible spatially unresolved outflow around BN, revealed as a distinct broad spectral component. \textit{Right:}
[\OI]\,63\,$\upmu$m spectra toward Source~I (red), BN (blue), and Peak~1 (black).
Only the blue spectrum  shows the broad spectral feature 
around the LSR velocity of BN  (dashed green line).}  
\label{fig:compact-outflow}
\end{figure*}

\subsubsection{OH P-Cygni profiles from the high-velocity outflow}

Figure~\ref{fig:spectra} (left panel, yellow histograms) 
display the \12OH~line profile (one of the $\Lambda$-doublets of the 
excited OH\,163\,$\upmu$m transition). These observations have
 much higher angular and spectral resolution than previous observations
with the Kuiper Airborne observatory (KAO),
the Infrared Space Telescope (ISO), and \textit{Herschel}
   \citep[][]{Melnick87,Melnick90,Goico06,Goipacs15}. The OH\,163\,$\upmu$m line profile  toward Peak~1 shows a very broad profile, with a FWZI of $\gtrsim$\,150\,km\,s$^{-1}$ and  $\Delta v_{\rm FWHM}$\,$\simeq$\,40\,km\,s$^{-1}$,   thus comparable to the CO
lines \citep[e.g.,][]{Kwan76,Snell84,Zapata09,Peng12,Bally17}.
As opposed to the atomic fine-structure lines,  the OH\,163\,$\upmu$m emission does not show a narrow 
line component. Indeed, the line is very faint toward the Trapezium position
(\mbox{Fig.~\ref{fig:spectra}}), implying that the broad-line emission originates in shocked gas and regions that are subject to strong radiative pumping, but not from the quiescent cloud gas.
This rotationally excited OH line arises from
the first excited state (at $E_{\rm u}/k$\,=\,270\,K) of the $^2\Pi_{1/2}$ ladder.
Given its high excitation requirements
 (\mbox{$n_{\rm cr}$\,$\simeq$\,10$^9$\,cm$^{-3}$}), the upper level can only be populated by collisions in very dense, warm gas, and/or by absorption of FIR dust photons in the \mbox{$^2\Pi_{1/2}$–$^2\Pi_{3/2}$} cross-ladder transition at $\sim$53\,$\upmu$m \mbox{\citep[e.g.,][]{Goico02,Goico06}}. Owing to the different dust continuum levels, the latter process contributes to the hot core region,
while the former likely dominates in  Peak~1 and Peak~2, with possible lateral
(i.e., not along the line of sight) FIR illumination from the hot core.
 The broad OH\,163\,$\upmu$m line profiles trace the outflow at all velocities, and the line intensity tends to peak toward the hot core 
(see also \mbox{Fig.~\ref{fig:channels}}), where the effect of FIR pumping is strongest.

Figure~\ref{fig:Figure_oh119} compares the  [\OI]\,63\,$\upmu$m, 
\mbox{CO~$J$\,=\,10--9}, and \32OH~(one of the $\Lambda$-doublets of the OH\,119\,$\upmu$m ground-state transition) line profiles toward the hot core position (roughly the FIR dust continuum peak).
The [\OI]\,63\,$\upmu$m spectrum shows the three absorption components  at $v_{\rm LSR}\simeq+6$, $+2$, and $-1$\,km\,s$^{-1}$.
The first component is consistent with gas in the hot core, while the other two originate
from foreground gas associated with Orion’s Veil.
\citet{Betz89} provided the first sub-km\,s$^{-1}$ resolution observations of the  OH\,119\,$\upmu$m rotational line  with the KAO. These observations revealed a self-absorption profile, characterized by absorption on the blueshifted side and emission on the redshifted side. The SOFIA
telescope delivers a factor of three better angular resolution (11.3$''$ at 119\,$\upmu$m), which provides less angular dilution of the   FIR continuum source. The resulting  OH\,119\,$\upmu$m line shows a more prominent \mbox{P-Cygni} profile (the line shows absorption and emission of similar strength), with a FWZI of $\simeq$\,160\,km\,s$^{-1}$, much like CO lines. Compared to \citet{Betz89}, SOFIA observations show that the entire blueshifted wing is absorbed from \mbox{$v_{\rm LSR} \simeq -80$} to $+8.5$\,km\,s$^{-1}$, while the redshifted emission extends up to \mbox{$v_{\rm LSR} \simeq +80$\,km\,s$^{-1}$}, with a velocity coverage similar to that of the
\mbox{CO $J$\,=\,10--9} line.
Clearly, OH is
present in the high-velocity outflowing gas, with terminal (maximum) velocities
of about  80\,km\,s$^{-1}$ --the fastest gas producing absorption projected along the line of sight. 
A closer examination of the OH\,119\,$\upmu$m P-Cygni profile reveals that the main absorption dip occurs at \mbox{$v_{\rm LSR} \simeq +6$ km s$^{-1}$}, which corresponds to the hot core absorption (for a zoomed-in spectrum, see \mbox{Fig.~\ref{fig:ZOOM_ALL_OH}}).
We leave a detailed analysis of the OH
excitation to a forthcoming paper.

\subsection{A possible outflow or wind bow shock around BN}\label{sec:compact-outflow}

\mbox{Figure~\ref{fig:compact-outflow} (left)} shows a map of the 
[\OI]\,63\,$\upmu$m emission in the $v_{\rm LSR}$ range from $+$20 to $+$25\,km\,s$^{-1}$.
In addition to the redshifted extended emission from the explosive outflow, we detect an emission peak near BN, located slightly northwest (NW) of this young massive star \citep[$\sim$\,8--13\,\Msun; e.g.,][]{Rodriguez05}.
This emission does not coincide with the more extended, so-called \mbox{low-velocity} \mbox{``18~km\,s$^{-1}$ outflow,''} which expands in the SW–NE direction \citep[][]{Genzel81,Plambeck82} and is likely driven by \mbox{Source I} \citep[][]{Beuther08,Wright22}.
The new [\OI]\,63\,$\upmu$m component arises from a broad emission feature, with
a line FWHM of \mbox{15\,$\pm$\,1\,km\,s$^{-1}$},  centered at
\mbox{$v_{\rm LSR}\approx$\,20\,km\,s$^{-1}$}
(blue spectrum in \mbox{Fig.~\ref{fig:compact-outflow}}).
This emission  is confined to the vicinity of BN, indicating either a compact outflow or a knot or shell of shock-excited gas surrounding the star.
It envelops the IR nebulosity associated with BN \mbox{\citep[see][]{Shuping04,Bally20}} but has not been reported in other line tracers, likely owing to the nature of this shocked gas and the complex substructure of the region in terms of emission, velocity, and extinction.

The centroid  of the [\OI]\,63\,$\upmu$m line feature is compatible with the LSR velocity measured from emission lines associated with BN
\citep[\mbox{$v_{\rm LSR} \simeq 23$~km~s$^{-1}$};][]{Gomez08,Goddi11,Plambeck13}.  
The apparent extent of this [\OI]\,63\,$\upmu$m  feature is
$<$\,10$''$ 
 and remains undetected at 145\,$\upmu$m.
The blue- and redshifted [\OI]\,63\,$\upmu$m emission around the LSR velocity of BN
shows a similar spatial distribution, suggesting that, if it originates from
an outflow, the outflow is spatially unresolved.
Given the uncertain properties of BN’s circumstellar disk, particularly its geometry \citep{Jiang05,Beuther10,Bally20}, it remains unclear whether this structure traces a disk-driven outflow or another form of shocked gas.

Interestingly, the [\OI]\,63\,$\upmu$m intensity peak is shifted to the NW of BN
by $\lesssim$\,3$''$, and is roughly aligned with the proper motion of this runaway star.
Thus, this [\OI]\,63\,$\upmu$m emission may correspond to an unresolved wind bow shock produced by the interaction of BN’s supersonic motion and wind with the ambient cloud gas \citep[][]{Baranov71,Wilkin96}. Adopting estimates of the wind mass-loss rate ($\dot{M_w}$), wind velocity ($v_w$), and BN’s velocity ($v_*$) relative to the surrounding cloud of density $n_0$ \citep[][]{Scoville83,Bally20}, the bow-shock radius ($r_{\rm bs}$) would  be several hundred au, or $\lesssim$\,1$''$ \citep{Tan04}. 
However, the wind mass-loss rate and velocity remain uncertain. Since $r_{\rm bs} \propto (\dot{M_w},v_w/n_0)^{1/2}\,v_{*}^{-1}$, larger wind parameters would yield a correspondingly larger $r_{\rm bs}$ (for fixed $n_0$ and $v_{*}$), potentially consistent with the apparent size of the [\OI]\,63\,$\mu$m feature.

\section{Analysis}\label{sec:analysis}

In the following, we restrict our analysis to a \mbox{100$''$ $\times$ 80$''$} (\mbox{0.2\,pc\,$\times$\,0.16\,pc}) region around the outflow, indicated with a dashed square in \mbox{Fig.~\ref{fig:sofia_maps_int_ratios}}. Specifically, we excluded the Trapezium region. In addition, we used the spectra of Peak~1 as a template to determine the physical conditions of the post-shock gas.

\subsection{Velocity-resolved line intensity ratios}
\label{subsec-line_profiles}
  
\begin{figure}[t]
\centering   
\includegraphics[scale=0.41,angle=0]{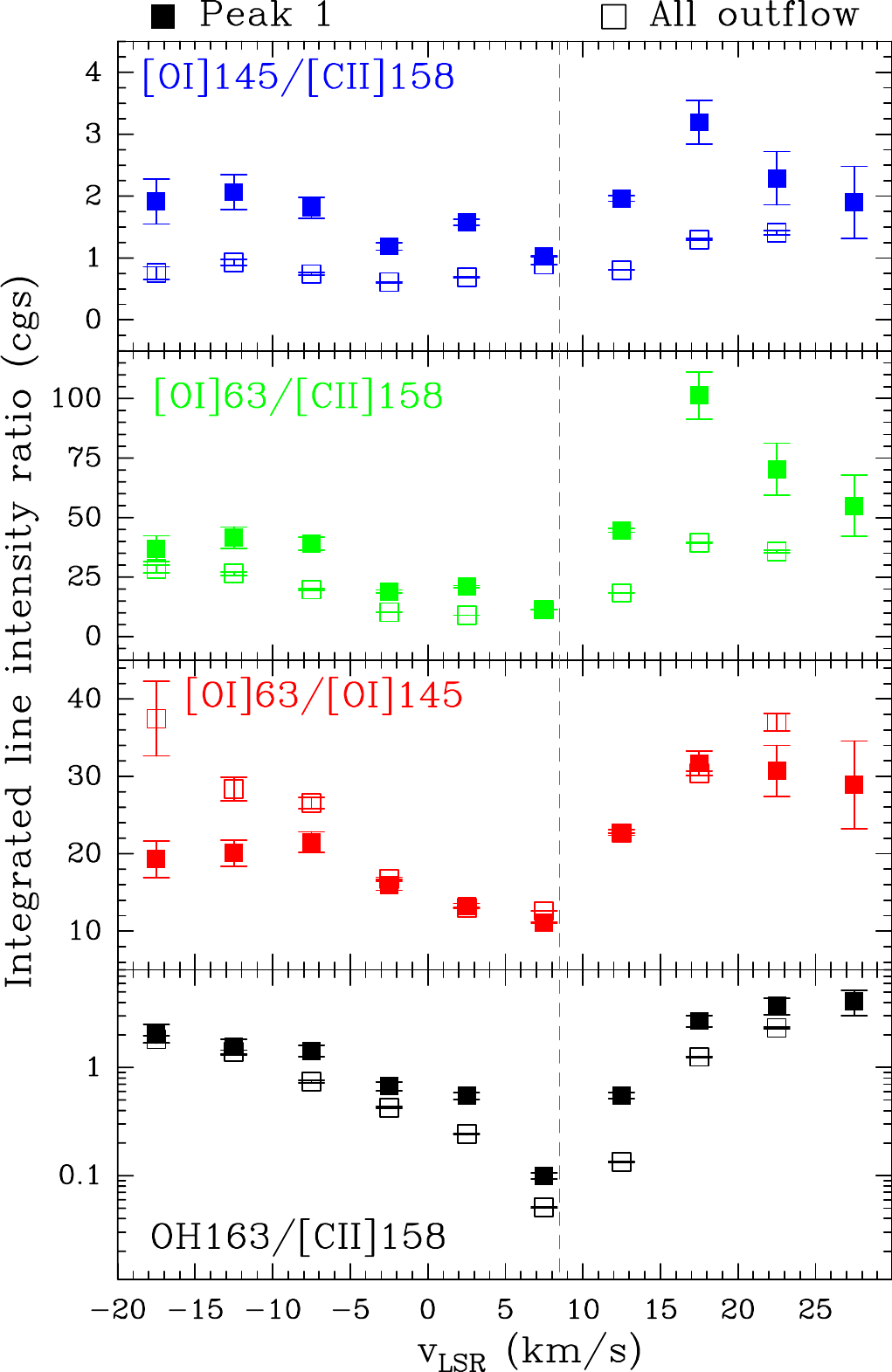}
\caption{Line intensity ratios as a function of velocity
in a 15$''$ beam  toward Peak~1 (filled squares)
and averaged over the \mbox{100$''$\,$\times$\,80$''$} region of the  BN/KL outflow  (empty squares), binned in intervals of 5\,km\,s$^{-1}$
from maps convolved to a common resolution of 15$''$. The dashed magenta line marks 
the LSR velocity of the quiescent gas in \mbox{OMC-1}.}  
\label{fig:wing_ratios}
\end{figure}

\mbox{Figure~\ref{fig:wing_ratios}} 
shows the FIR line intensity ratios as a function of velocity toward
the Peak~1 position (filled squares) obtained from maps convolved to a common angular resolution of 15$''$. The lowest [\OI]\,63/145 line intensity ratio, 11.4\,$\pm$\,0.1, occurs at $v_{\rm LSR,0}$, corresponding to the LSR velocity of the quiescent gas in \mbox{OMC-1} (dashed magenta line in \mbox{Fig.~\ref{fig:wing_ratios}}), 
where the narrow line component is dominated by emission from the extended PDR in front of the outflow. This intensity ratio is comparable to that observed in the Orion Bar \citep[e.g.,][]{Salas12}. The [\OI]\,63/145 intensity ratio then increases in the line wings, reaching $\sim$30  at redshifted velocities.
The [\OI]\,63/145 intensity ratio shows an asymmetry between
the blue- and redshifted wings. 
Foreground [\OI]\,63\,$\upmu$m absorption at $v_{\rm LSR}$ between roughly $-5$ and $+7$\,km\,s$^{-1}$ contributes to this asymmetry, but the blueshifted wing shows smaller ratios at higher, more negative velocities.
These lower  ratios suggest a correspondingly lower density in the blueshifted post-shock  gas (\mbox{Sect.~\ref{sub-sec:MTC-models}}).

\mbox{Figure~\ref{fig:wing_ratios}} also shows the velocity dependence of the \mbox{[\OI]\,145\,/\,[\CII]\,158}, 
\mbox{[\OI]\,63\,/\,[\CII]\,158}, and \mbox{OH\,163\,/\,[\CII]\,158} line intensity ratios.
All these ratios reach a minimum at the LSR velocity of the 
 \mbox{face-on} PDR, $v_{\rm LSR,0}$,  and gradually increase in the line  wings, indicating a lower contribution of C$^+$ in the highest velocity gas.
The [\OI]\,145\,$\upmu$m line  is particularly bright, with \mbox{[\OI]\,145\,/\,[\CII]\,158} intensity ratios of \mbox{$\simeq$\,2--3} in the line wings.
 The \mbox{OH\,163/[\CII]\,158} line intensity ratio increases from $\simeq$\,0.1 at the line center to 4\,$\pm$\,1 in the wings, thus showing a strong dependence on radial velocity.
The redshifted line wing shows very high  \mbox{[\OI]\,63\,/\,[\CII]\,158}  intensity ratios, up to $\simeq$50–100, much higher than the expected ratios in PDRs
with $n_{\rm H} \approx 10^5$\,cm$^{-3}$ \citep[e.g.,][]{Kaufman99}.
For reference, in the prototypical dense PDR, the Orion Bar, at  $\sim$\,2$'$ south of the Trapezium,
these line intensity ratios are much lower, \mbox{[\OI]\,63/145\,$\simeq$\,9}, \mbox{[\OI]\,63\,/\,[\CII]\,158\,$\simeq$\,8}, \mbox{[\OI]\,145\,/\,[\CII]\,158\,$\simeq$\,0.9},
and \mbox{OH\,163\,/\,[\CII]\,158\,$\simeq$\,0.03} \mbox{\citep[e.g.,][]{Salas12,Joblin18}}.

\begin{table*}[!h] 
\begin{center} 
\caption{FIR atomic fine-structure line intensities and intensity ratios at different velocities toward Peak 1 and the Orion Bar PDR (for reference).} 
\label{Table_intensities_wings}  
\normalsize
\begin{tabular}{r c c c c c c @{\vrule height 10pt depth 5pt width 0pt}}    
\hline\hline       
\hspace{1.4cm} [$v_{\rm LSR}$ range]     & [\OI]\,63\,$\upmu$m                    &          [\OI]\,145\,$\upmu$m           &  [\CII]\,158\,$\upmu$m                 &                                         &                                &   \\
  Source\,\,\,\,\,\,\,\, (km\,s$^{-1}$)  &  ($\times$10$^{-2}$ erg\,s$^{-1}$\,cm$^{-2}$\,sr$^{-1}$) &    & \hspace{-2.cm}($\times$10$^{-4}$ erg\,s$^{-1}$\,cm$^{-2}$\,sr$^{-1}$)   & $I_{63}$/$I_{145}$    & $I_{63}$/$I_{158}$ & $I_{145}$/$I_{158}$ \\\hline
Red wing: $[+21,+46]$     &             1.0$\pm$0.1                        &                             3.7$\pm$0.4              &      1.7$\pm$0.3                                                        &  27$\pm$4  & 59$\pm$10 & 2.2$\pm$0.4 \\
Blue wing: $[-30,-5]$     &             1.8$\pm$0.1                        &                             9.8$\pm$0.4              &      4.9$\pm$0.3                                                        &  18$\pm$1  & 37$\pm$3 & 2.0$\pm$0.1\\
Narrow: $[+5,+10]$        &             4.0$\pm$0.1                        &                             36.0$\pm$0.2             &      35.0$\pm$0.1                                               &  11$\pm$1  & 11$\pm$1 & 1.0$\pm$0.1\\\hline
Orion Bar PDR$\ddagger$   &             19$\pm$5                   &                            210$\pm$40                 &       230$\pm$90                                        &  9$\pm$3   & 8$\pm$4 & 0.9$\pm$0.4\\\hline
\end{tabular}                                                                                             
\end{center} 
\normalsize
\vspace{-0.5cm}
\tablefoot{In a beam of 15$''$ from maps convolved to a common resolution of 15$''$.
$\ddagger$From \textit{Herschel} \citep{Salas12,Joblin18}.}                
\end{table*}      

The empty squares in \mbox{Fig.~\ref{fig:wing_ratios}} show the FIR line intensity ratios as a function of LSR velocity, averaged over the \mbox{100$''$\,$\times$\,80$''$} region, thus representing the average conditions in the  outflow. These ratios are similar to those toward Peak~1, with the main differences being a slightly higher [\OI]\,63/145 intensity ratio in the blueshifted wing and slightly lower \mbox{[\OI]\,145/[\CII]\,158}, \mbox{[\OI]\,63/[\CII]\,158}, and \mbox{OH\,163/[\CII]\,158} ratios. This reflects the higher relative contribution of the [\CII]\,158\,$\upmu$m emission\footnote{The [\CII]\,158\,$\upmu$m emission around 
$v_{\rm LSR}$\,$\simeq$\,$+$20\,km\,s$^{-1}$ can be dominated by
[$^{13}$\CII]\,\mbox{$F$\,=\,2--1} emission from \mbox{OMC-1} (right panels of Fig.~\ref{fig:spectra}).} in the SE part of the outflow (see Figs.~\ref{fig:sofia_maps_int_ratios}b and \ref{fig:sofia_maps_int_ratios}c).
\mbox{Figure~\ref{fig:channels_ratios}} shows these results in the form of intensity-ratio velocity channel maps. 
These maps illustrate the same trends described above: the \mbox{[\OI]\,63/[\CII]\,158} and \mbox{[\OI]\,145/[\CII]\,158} ratios in the redshifted wing peak toward Peaks~1 and~2, following the H$_2$ outflow.  These  regions coincide with enhanced \mbox{[\OI]\,63/145} intensity ratios.

 \mbox{Table~\ref{Table_intensities_wings}} summarizes the intensities of the atomic fine-structure lines toward Peak~1, measured in three velocity intervals of the line profile: the redshifted wing, the blueshifted wing, and the narrow component. To increase the S/N of the line-intensity measurements and to cover the relevant outflow velocity range, we derived the wing intensities using velocity bins of 25\,km\,s$^{-1}$ in the wings. We adopted the 
 $[+21, +46]$ and $[-30, -5]$\,km\,s$^{-1}$ LSR velocity intervals for the redshifted and blueshifted wings, respectively.
 That is, radial velocities of 13 to 38 km\,s$^{-1}$ with respect to $v_{\rm LSR,\,0}$.
  This choice excludes the velocity ranges of the [\OI]\,63 and [\CII]\,158\,$\mu$m lines that are affected by foreground absorption and emission in the blueshifted wing (see Fig.~\ref{fig:spectra}). We also subtracted the contribution of the [$^{13}$\CII] \mbox{$F$\,=\,2--1} hyperfine-structure line in the redshifted wing using multi-Gaussian fitting.
In \mbox{Sect.~\ref{subsec:shock-models}}, we  use these line-wing intensities to constrain the nature of the shocks that give rise to the FIR atomic line emission.

\subsection{Post-shock gas physical conditions and   [\OI]63,145\,$\mu$m excitation  across different  outflow velocity components} \label{sub-sec:MTC-models}

The observed \mbox{[\OI]\,63\,/\,[\OI]\,145\,$\upmu$m} intensity ratios (\mbox{$I_{63}/I_{145}$}) in the line wings are high, reaching up to 
$\sim$\,30 (e.g., \mbox{Table~\ref{Table_intensities_wings}}). 
This is close to the high-temperature limit for optically thin emission
under local thermodynamic equilibrium (LTE\footnote{In the optically thin LTE regime, \mbox{$I_{63}/I_{145} = 35.3\,e^{+98.8/T_{\rm k}}$},
where  \mbox{[$E$($^3$P$_0$)$-$$E$($^3$P$_1$)]\,/\,$k$\,=\,98.8\,K \citep[see also][]{Goldsmith19}.}}) conditions (\mbox{$T_{\rm ex}$\,$=$\,$T_{\rm k}$}). 
This would imply $N$(O)\,$<$ a few \mbox{10$^{18}$\,cm$^{-2}$} and high  densities, meaning
\mbox{$n_{\rm H}\gg n_{\rm cr}$}, with \mbox{$n_{\rm cr,\,63\upmu m}\simeq 5 \times 10^5$\,cm$^{-3}$}
and \mbox{$n_{\rm cr,\,145\upmu m}\simeq 6 \times 10^6$\,cm$^{-3}$} for optically thin emission. However, LTE is rarely fulfilled, and even for a three-level system such as the $^3$P$_J$ levels of oxygen, subtle non-LTE (NLTE), line opacity, and FIR pumping effects can dominate the [\OI] emission,
thus, the resulting $I_{63}/I_{145}$ ratio \citep[][]{Monteiro87,Elitzur06,Goico09,Goldsmith19}.

To estimate the beam-averaged physical conditions ($n_{\rm H}$, $T_{\rm k}$) and the atomic oxygen column densities, $N$(O), in the post-shock gas across the different velocity components, we used a nonlocal NLTE radiative transfer model \citep{Goico06b,Goico09} with up-to-date \mbox{O--H$_2$} and \mbox{O--H} inelastic collision rate coefficients 
\citep{Lique18}. This \textit{nonlocal} approach accounts for [\OI]\,63\,$\upmu$m line opacity and line trapping, as well as [\OI]\,145\,$\upmu$m suprathermal excitation (\mbox{$T_{\rm ex} > T_{\rm k}$}) and population inversions, which "local" single escape probability methods fail to correctly capture. 
To match the observed [\OI]\,145\,$\upmu$m line FWHM (broad component), we adopted a turbulent line
width of \mbox{20\,km\,s$^{-1}$}.  
\mbox{Following} \citet{Goico09}, we also examined the impact of FIR radiative pumping on the [\OI] fine-structure level populations.
\mbox{Appendix~\ref{App:MTC_models}} provides  details  and explores the excitation conditions over a wide parameter space.

\begin{figure}[t]
\centering    
\includegraphics[scale=0.53,angle=0]{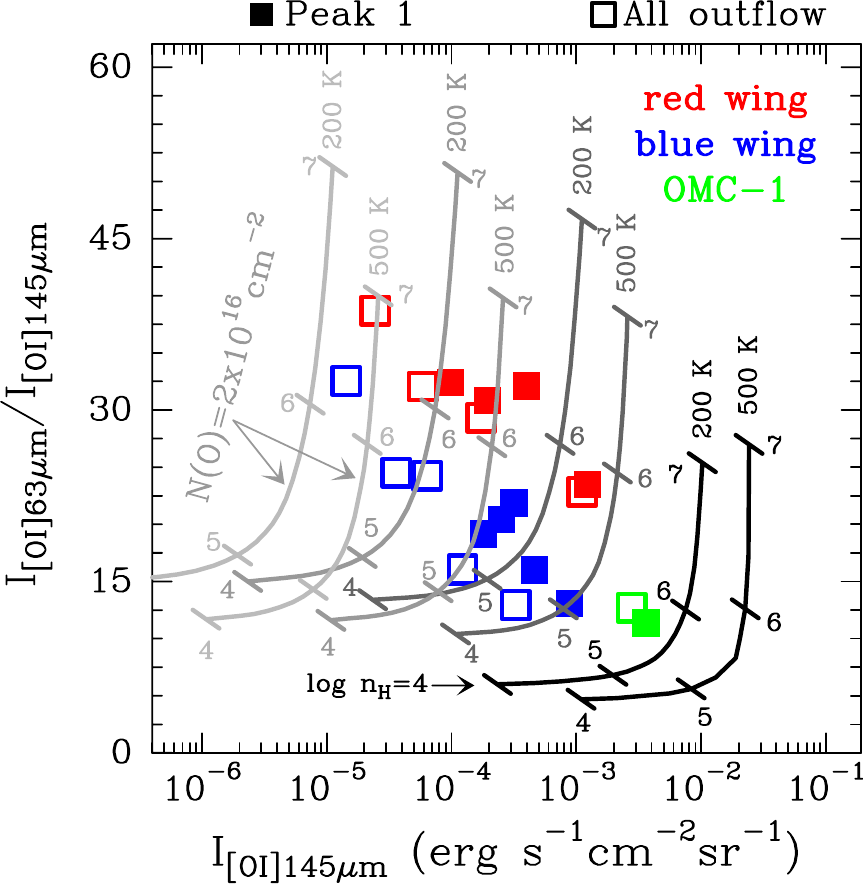}
\caption{[\OI]\,63\,/\,145\,$\upmu$m  intensity ratio versus  [\OI]\,145\,$\upmu$m  intensity. The light grey to black curves show a grid of  radiative transfer models for four 
values of $N$(O): \mbox{2$\times$10$^{16}$}, \mbox{2$\times$10$^{17}$}, \mbox{2$\times$10$^{18}$}, and \mbox{2$\times$10$^{19}$\,cm$^{-2}$}, from left to right, at 200 and 500\,K.
Tick marks denote log\,$n_{\rm H}$ (in cm$^{-3}$) and are separated by one decade along curves of constant $N$(O).
The colored squares show the observed values in LSR velocity bins of 5\,km\,s$^{-1}$ toward Peak~1 (filled squares) and averaged over the  BN/KL outflow (open squares). Red squares correspond to the redshifted line wing,
 while blue squares correspond to the blueshifted line wing. 
Green squares correspond to  \mbox{OMC-1} velocities (see  Tables~\ref{Table_intensities_vPeak1} and \ref{Table_intensities_voutflow}).}  \label{fig:model_MTC_outflow}
\end{figure}

Figure~\ref{fig:model_MTC_outflow} presents the results of a model grid
displayed in the velocity-dependent parameter space \mbox{$I_{63}/I_{145}\,(v)$ versus $I_{145}(v)$}.
The light grey to black curves show radiative transfer models for four atomic oxygen column densities, from \mbox{2$\times$10$^{16}$} to  \mbox{2$\times$10$^{19}$\,cm$^{-2}$}, increasing from left to right, at two gas temperatures, 200 and 500\,K.
Tick marks denote log\,$n_{\rm H}$ (in cm$^{-3}$) and are separated by one decade along curves of constant $N$(O).
 In general,
as the opacity of the [\OI]\,63\,$\upmu$m line increases, the intensity ratio \mbox{$I_{63}/I_{145}$}  decreases roughly as \mbox{$\sim 1/\tau_{63}$}, while the [\OI]\,145\,$\upmu$m line is rarely optically thick.
In dense gas, intensity ratios \mbox{$I_{63}/I_{145}$\,$\lesssim$\,15} typically imply optically thick [\OI]\,63\,$\upmu$m  emission.
For the adopted parameters, the gas density also determines the 
[\OI] excitation regime.
For \mbox{$n_{\rm H} \gtrsim n_{\rm crit}$}, the \mbox{$I_{63}/I_{145}$} ratio increases with $n_{\rm H}$ owing to a significant increase in the [\OI]\,63\,$\upmu$m excitation temperature and a decrease in $\tau_{63}$ toward the optically thin regime. In this regime, the ratio also increases with decreasing gas temperature.
The colored squares in \mbox{Fig.~\ref{fig:model_MTC_outflow}} represent the observed values in LSR velocity bins of 5\,km\,s$^{-1}$ toward Peak~1 (filled squares) and averaged over the  BN/KL outflow (open squares). Red squares correspond the redshifted line wing, in the range $[+10,+25]$~km\,s$^{-1}$, while blue squares correspond to the blueshifted line wing, $[-20,0]$~km\,s$^{-1}$.
Green squares in \mbox{Fig.~\ref{fig:model_MTC_outflow}}  correspond to the quiescent cloud gas at \mbox{OMC-1} velocities, \mbox{$v_{\rm LSR}$\,=\,$[+5,+10]$~km\,s$^{-1}$}. The green squares are consistent with large column densities
of  atomic oxygen, 
\mbox{$N$(O)\,$\simeq$\,10$^{19}$\,cm$^{-2}$}, optically thick [\OI]\,63\,$\upmu$m  emission, $n_{\rm H}$ of several 10$^5$\,cm$^{-3}$, and  temperatures 
\mbox{$T_{\rm k}$\,$\simeq$\,200--400\,K}--thus compatible with
the observed $T_{\rm P}$([\OI]63\,$\upmu$m) values (\mbox{Sect.~\ref{sub-sec:profiles}})--, typical of a dense
PDR with a thermal pressure, $P_{\rm th}/k$, of 10$^7$ to 10$^8$\,K\,cm$^{-3}$.
This agrees with previous observations of the large-scale PDRs at the surfaces
of \mbox{OMC-1}
\citep[][]{Herrmann97,Goicoechea15,Goico19}.
An increase in velocity within the line wings (red and blue squares)  is accompanied by a rise in the [\OI]\,63/145 intensity ratio, consistent with higher gas densities and less opaque [\OI]\,63\,$\upmu$m emission. The lower $I_{145}(v)$
intensities as velocities increase simply reflect smaller $N$(O) column densities. 
Toward Peak~1, the 
[\OI]\,63 and 145\,$\upmu$m line-wing emission is
consistent with $n_{\rm H}$ of several 10$^5$ to \mbox{10$^6$\,cm$^{-3}$}, 
\mbox{$T_{\rm k}$\,$\lesssim$\,500\,K}, and $N$(O) of a few 10$^{17}$\,cm$^{-2}$.
\mbox{Figure~\ref{fig:model_MTC_outflow}} also indicates that the post-shock gas density 
is slightly lower in the blueshifted gas.
When averaged over the entire outflow, the \mbox{$I_{63}/I_{145}(v)$} ratio is highest at the largest velocities, where the H$_2$ and O column densities are lowest. The higher ratios imply
$n_{\rm H}$ up to several $10^6$\,cm$^{-3}$, consistent with gas compression
 in the post-shock gas.

\section{Discussion}\label{sec:discussion}

\subsection{Peak~1 as a laboratory for shock modeling}

Early models of Peak~1  have  required  two types of shocks to explain the observations \cite[e.g.,][]{Chernoff82,Draine82,Neufeld89b,Haas91,Kaufman96,LeBourlot02,Esplugues14}.
The accepted picture was that a high-velocity molecular flow (the “plateau”) produces nondissociative \mbox{$C$-type} shocks (with shock velocities \mbox{$v_S \gtrsim 25$\,km\,s$^{-1}$}) as the flow interacts with the ambient molecular cloud, generating high temperatures (above 1000\,K). 
These shocks lead to bright H$_2$ and high-$J$~CO emission along with large \mbox{H$_2$O/CO\,$>$\,1} abundance ratios. 
Newer \mbox{$C$-type} shock models included the effects of modest external UV illumination \citep{Melnick15} to explain the unusual detection  of narrow-line O$_2$ toward Peak~1 \citep{Chen14}. 
In addition, fast, \mbox{$v_S$\,=\,70--80\,km\,s$^{-1}$}, dissociative, \mbox{$J$-type} shocks were proposed to occur where a high-velocity wind impacts the expanding swept-up material \citep[][]{Hollenbach85}, producing bright atomic fine-structure line emission. However, BN/KL's explosive outflow is driven by bullets moving through the molecular cloud \citep[e.g.,][]{Bally15,Bally20}, rather than by stellar jets or winds, as in protostellar outflows. \mbox{In Sect.~\ref{subsec:shock-models}} we explore the nature
of the shocks driving the atomic fine-structure emission.

\subsection{Nature of the [\OI]--emitting shocks}\label{subsec:shock-models}

\begin{figure}[t]
\centering   
\includegraphics[scale=0.435,angle=0]{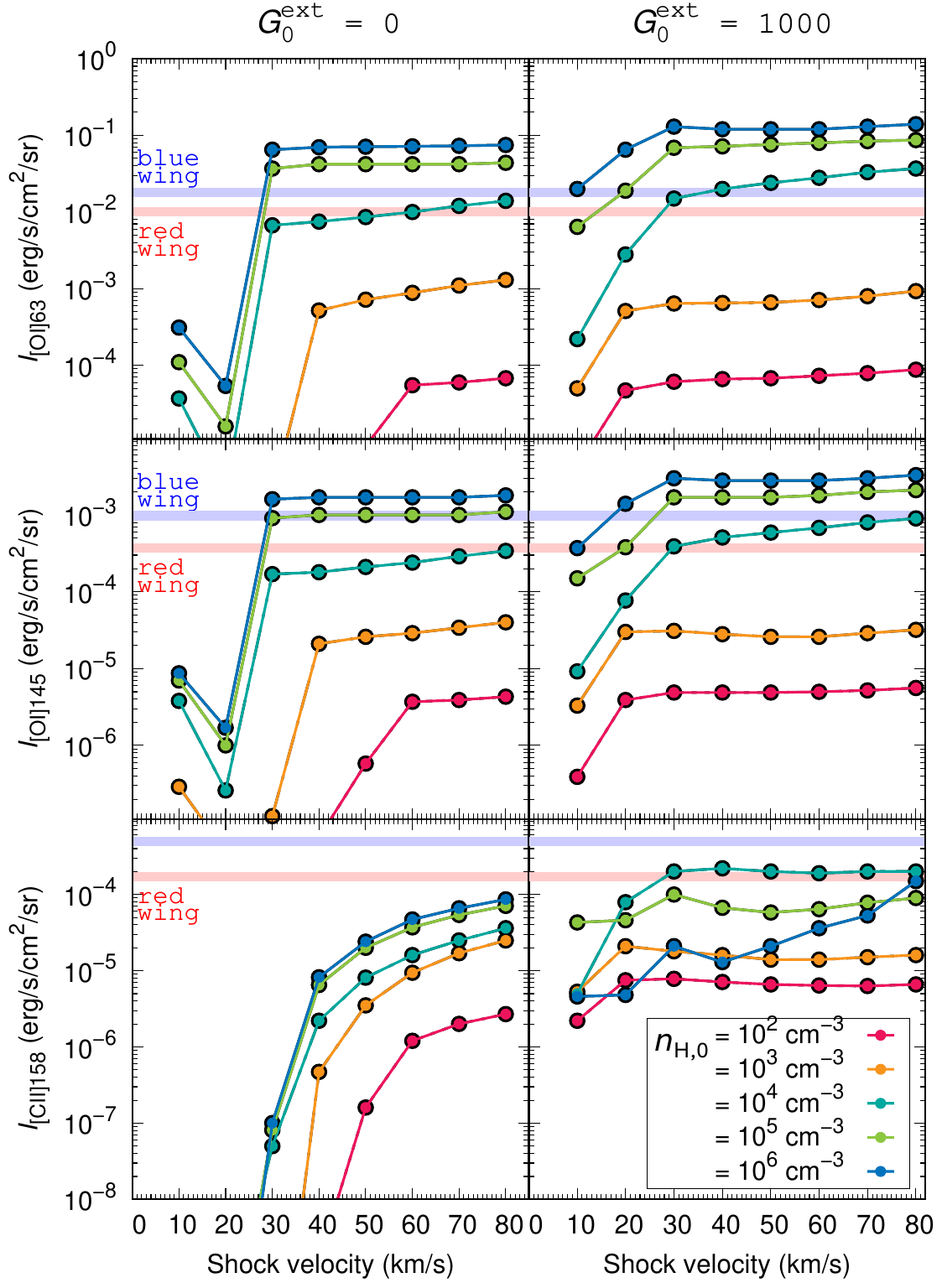}
\caption{FIR atomic line intensities as a function of shock velocity and preshock gas density \citep[from][]{Lehmann22}. 
The right panels show externally irradiated shocks. The horizontal blue and red lines indicate the observed line-wing intensities toward Peak~1 (Table~\ref{Table_intensities_wings}).} 
\label{fig:shock_models}
\end{figure}

Figure \ref{fig:shock_models} compares the intensities of the 
[\OI]\,63\,$\upmu$m, [\OI]\,145\,$\upmu$m, and [\CII]\,158\,$\upmu$m lines observed in the red and blue wings toward Peak~1 (Table~\ref{Table_intensities_wings}) with those predicted by the Paris–Durham shock code\footnote{Available on the ISM platform \url{https://ism.obspm.fr}.} \citep{Godard19} along the direction perpendicular (face-on) to the shock front. 
In this comparison, we assume an emission  beam filling factor of $f=1$. 
The models are from \citet{Lehmann22}, which account for magnetized molecular shocks with velocities in the range 
\mbox{$v_S=5$–$80\,\mathrm{km\,s^{-1}}$}, preshock gas densities in the range \mbox{$n_{\rm H,0}=10^{2}$–$10^{6}\,\mathrm{cm^{-3}}$}, enabling us to include an external FUV  field ($G_{0}^{\rm ext}$), in units of the Mathis  field \mbox{\citep{Mathis83}}.

The strength of the preshock magnetic field  prior to the explosion is uncertain. Dust polarimetric observations of the plane-of-sky magnetic field in the quiescent cloud indicate values of a few hundred $\upmu$G \citep[][]{Houde09,Guerra21}, consistent with line-of-sight field strengths derived from CN Zeeman measurements \citep[][]{Crutcher99,Crutcher99b}. Thus, all models in \mbox{Fig.~\ref{fig:shock_models}} adopt a reference magnetic parameter \mbox{$b=1$}, where the preshock transverse magnetic field strength is given by \mbox{$B = b\,(n_{\rm H,0}/{\rm cm}^{-3})^{1/2}$\,$\upmu$G}.  
 For this range of parameters,  shocks can be of type \mbox{$C$}, $C^{*}$, or $CJ$ at \mbox{$v_S$\,$<$\,25\,km\,$^{-1}$}, or \mbox{$J$-type} at \mbox{$v_S$\,$>$\,25\,km\,$^{-1}$} \citep[][]{Lehmann20,Lehmann22}. 
 The models assume either \mbox{$G_{0}^{\rm ext}=0$} (left panels) or \mbox{$G_{0}^{\rm ext}=10^{3}$} (right panels). Accurately determining the flux of external FUV photons reaching each bow-shock structure—whether locally self-generated, generated in nearby shocks, emitted by massive protostars within the region \citep[BN, \mbox{source I}, and \mbox{source n};][]{Shuping04}, or originating from the Trapezium stars—is challenging, as it depends on both the porosity of the medium and the properties of the shocks and the protostars. The representative value of \mbox{$G_{0}^{\rm ext} = 10^{3}$} lies within the very rough estimates from FIR continuum observations toward Peak~1 \citep[][]{Goipacs15}.
 
In the absence of external irradiation (\mbox{$G_{0}^{\rm ext}=0$}), the predicted [\OI] line intensities converge toward a plateau at high shock velocities, the level of which is primarily set by the preshock gas density. This is
consistent with the dissociative shock models of
\citet{Hollenbach89}.
Shocks with sufficiently high velocities generate a local FUV radiation field \mbox{\citep[e.g.,][]{Lehmann20}}, whose intensity 
scales as
\begin{equation}
\label{eq-UV-shocks}
G_0^{\rm shock} = 10^3 \left( \frac{n_{\rm H,0}}{10^4\,{\rm cm}^{-3}} \right) \left( \frac{v_S}{80\,{\rm km~s}^{-1}}\right)^{3},
\end{equation} 
for  \mbox{$30 < v_S \lesssim 100$\,km\,s$^{-1}$} \citep[][]{Lehmann20,Godard24},
 and thus strongly depend \mbox{on $v_S$}.
At low shock velocities, models with \mbox{$G_{0}^{\rm ext}$\,=\,0} 
underpredict the observed 
\mbox{[\CII]\,158\,$\upmu$m} intensity by several orders of magnitude.
Marginally reproducing the [\CII]158\,$\upmu$m intensity would require very fast shocks, \mbox{$v_S \gtrsim 80$ km s$^{-1}$}. \mbox{However},  this is
in contradiction with the  $\Delta v_{\rm FWHM}$ and FWZI values
of the observed fine-structure lines  (\mbox{Sect.~\ref{sub-sec:profiles}}), assuming that 
the flow is roughly directed along the line of sight.
Therefore, a significant contribution from locally self-generated UV radiation is unlikely.

External FUV radiation, which leads to  carbon photoionization and CO photodissociation, significantly enhances the intensities of atomic fine-structure lines
(right panels of \mbox{Fig.~\ref{fig:shock_models}}).
Still, owing to the enhanced importance of C$^+$ recombination and chemical destruction at high gas densities, the \mbox{[\CII]\,158\,$\upmu$m} intensity does not increase monotonically with density.

Under external irradiation, the observations no longer constrain a unique set of shock parameters, but are instead consistent with a family of solutions involving \mbox{$J$-type}  shocks with 
\mbox{$v_S$\,$\simeq$\,30--40\,km\,s$^{-1}$} and \mbox{$n_{\rm H,0}$$\,\gtrsim$\, a few $10^{4}$\,cm$^{-3}$}. 
Models with $G_{0}^{\rm ext}=10^{3}$ reproduce the [\CII]\,158\,$\upmu$m  intensity of the redshifted wing, tracing gas moving into \mbox{OMC-1}, whereas the blueshifted wing emission, tracing gas moving toward the Trapezium, appears to require  higher values of $G_{0}^{\rm ext}$. In the following, we focus on the redshifted wing emission. 

\begin{figure}[t]
\centering   
\includegraphics[scale=0.31,angle=0]{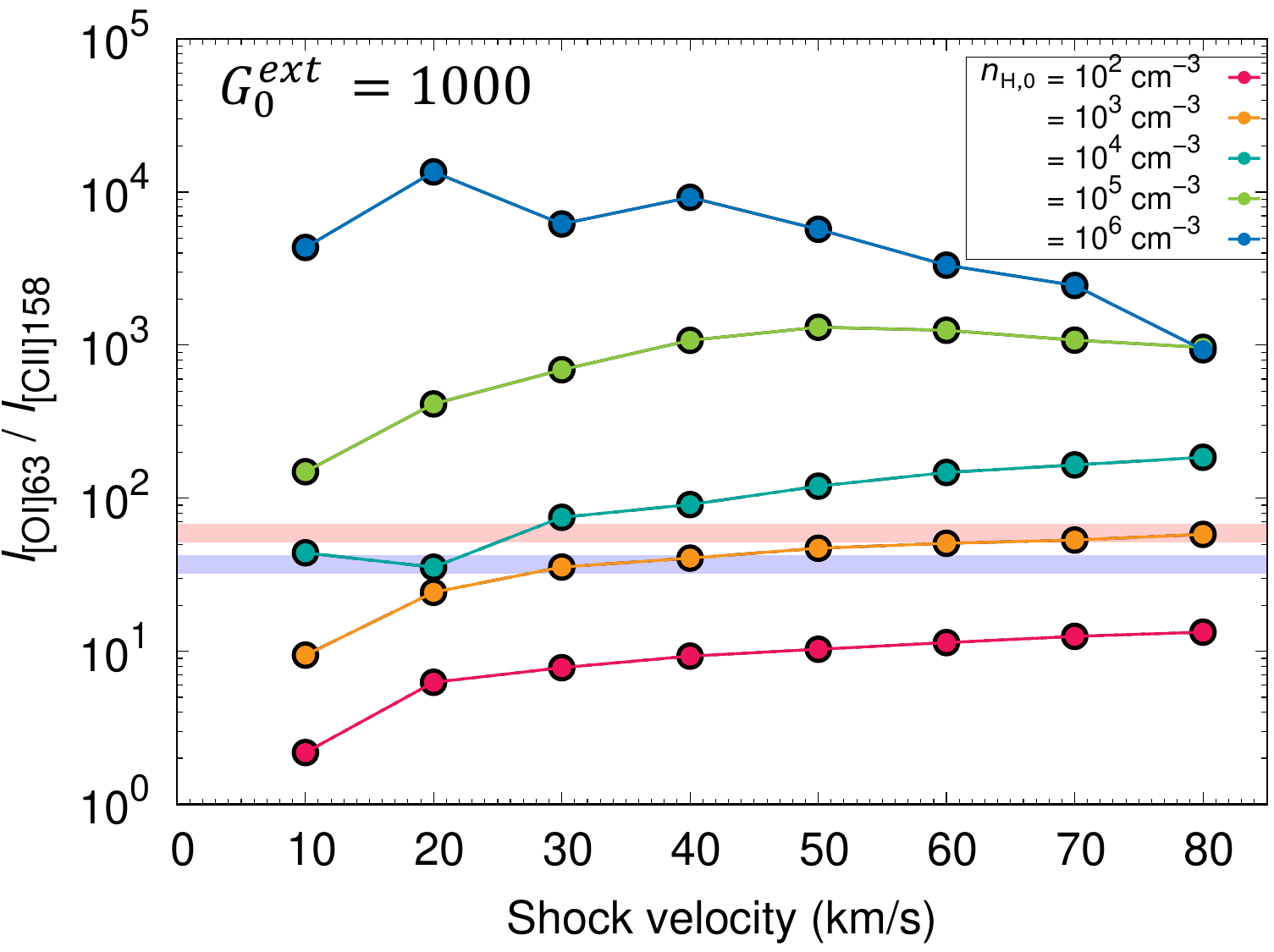}
\caption{[\OI]\,63\,/\,[\CII]\,158\,$\upmu$m line intensity ratio
 as a function of shock velocity and preshock gas density 
 in an externally irradiated shock with $G_{0}^{\rm ext}$\,=\,10$^3$ \citep[from][]{Lehmann22}. The horizontal blue and red lines indicate the observed line-wing intensities toward Peak~1.} 
\label{fig:ratio_shock_models}
\end{figure}

In these irradiated shocks, the predicted \mbox{$I_{63}/I_{158}$} line intensity ratio primarily scales with the preshock gas density and shows only a weak dependence on the shock velocity for \mbox{$v_S > 30$\,km\,s$^{-1}$} (see \mbox{Fig.~\ref{fig:ratio_shock_models}}). The above family
of solutions predicts \mbox{$I_{63}/I_{158}\,\gtrsim\,60$}, consistent with the redshifted wing emission toward Peak~1, which shows intensity ratios up to $\sim$50--100 
(\mbox{Fig.~\ref{fig:wing_ratios}}).
Given the inferred post-shock densities from the [\OI] excitation analysis in \mbox{Sect.~\ref{sub-sec:MTC-models}}, $n_{\rm H}$ of several $10^5$ to $10^6$\,cm$^{-3}$,  the resulting compression factors, \mbox{$F = n_{\rm H}/n_{\rm H,0}\,\gtrsim\,30$}, also support magnetized \mbox{$J$-type} shocks,
where \mbox{$F$\,$\sim$\,$v_S/b$} \citep{Godard24}.
The compressed magnetic field would have a strength of several mG, similar
to the values observed toward BN/KL \citep[][]{Pattle17,Guerra21}.
All in all, the  [\OI] and [\CII] line-wing emission is best reproduced by externally irradiated, dissociative \mbox{$J$-type} shocks with velocities of \mbox{30--40\,km\,s$^{-1}$}, preshock densities of a few $10^{4}$\,cm$^{-3}$, and compression factors consistent with post-shock densities derived from the  [\OI] excitation analysis.

 A more detailed interpretation would require the geometry, time dependence,
and spatial variations of $b$ and $G_{0}^{\rm ext}$ across the outflow to be fully accounted for. 
Still, the results shown in \mbox{Fig.~\ref{fig:shock_models}} provide a general interpretative framework: dense bullets, with proper motions up to several hundred \mbox{km\,s$^{-1}$},  accelerated by the explosion event \mbox{\citep[i.e.,][]{Doi02,Bally15,Bally17}}, interact with the ambient molecular cloud, generating bow-shock structures seen in [\FeII] knots
and H$_2$ (\mbox{Fig.~\ref{fig:H2}}). The highest bullet velocities exceed the observed FWZIs of the CO and OH lines, and imply molecular gas dissociation and the production of significant FUV radiation in many bow-shock tips (Eq.~\ref{eq-UV-shocks}). 
Within this framework, surrounding fast shocks generate sufficient external FUV radiation to illuminate the inner layers of the bow-shock structures. The working surfaces of these bow shocks drive dissociative \mbox{$J$-type} shocks with velocities of a few tens of km\,s$^{-1}$, producing the observed [\OI] emission.

Consistently with this scenario, the observed \HI~21\,cm line emission from atomic hydrogen (produced after H$_2$ (photo)dissociation) extends only up to \mbox{$v_{\rm LSR}\leq31$\,km\,s$^{-1}$} \citep[\mbox{see \mbox{Fig.~20} of}][]{vanderWerf13} and broadly resembles the [\OI]\,63\,$\upmu$m emission, both in its spatial distribution and in its lack of the highest velocity emission seen in CO or OH.
The broader CO and OH line profiles indicate that the shocks responsible for their emission differ from those driving the atomic fine-structure emission.
For instance, \citet{Melnick15} concluded that a \mbox{$C$-shock} with $v_S = 23\,\mathrm{km\,s^{-1}}$ propagating into a more shielded molecular clump (preshock density of $8\times10^4\,\mathrm{cm^{-3}}$ and $G_{0}^{\rm ext}=1$) explains  the  detection of O$_2$   \citep{Chen14}.

We conclude that, in contrast to what was found using previous models \citep[e.g.,][]{Hollenbach89,Haas91}, fast shocks with \mbox{$v_S \simeq 70$--$80\,\rm km,s^{-1}$} are not required to explain the observed atomic fine-structure line intensities and profiles. Instead, the observed lines are consistent with lower velocity, dissociative \mbox{$J$-type} shocks illuminated by external FUV radiation, with \mbox{$G_{0}^{\rm ext} \simeq 10^2$--$10^3$}, depending on the shock parameters \citep[e.g.,][]{Kristensen23}. This radiation arises from surrounding fast shocks and possibly from nearby massive (proto)stars, enhancing the [\CII]\,158\,$\upmu$m  emission, particularly in blueshifted gas moving toward the cluster.

\begin{table}[t] 
\begin{center}
\caption{Total IR line luminosities in the BN/KL outflow region$\dagger$.} 
\label{Table_intensities}  
\small
\normalsize 
\begin{tabular}{l c  @{\vrule height 10pt depth 5pt width 0pt}}    
\hline\hline       
Line                       & Luminosity (\Lsun) \\\hline
$[$\OI$]$\,63\,$\upmu$m    & 81.3$\pm$4.1$^{a,b}$       \\
$[$\OI$]$\,145\,$\upmu$m   & 5.2$\pm$0.3$^{a,b}$    \\
$[$\CII$]$\,158\,$\upmu$m  & 5.8$\pm$0.3$^{a,b}$      \\
H$_2$O (FIR and MIR)       & 35$\pm$2$^c$         \\     
OH (FIR)                   &   13$\pm$1$^c$         \\
CO (FIR and MIR)           & $\sim$98$\pm$9$^c$   \\
H$_2$ all lines            & $\sim$120$\pm$60$^d$  \\\hline
Total                      & $\sim$350$^a$ ($\gtrsim$\,320 from the outflow$^b$ itself)\\\hline
\end{tabular}                                                                                             
\end{center} 
\normalsize
\vspace{-0.5cm}
\tablefoot{$\dagger$Integrated  over each full line profile. 
$^a$Extracted over a \mbox{100\,$''\times$80\,$''$} area  
 (this work, box in Fig.~\ref{fig:sofia_maps_int_ratios}).
$^b$In Sect.~\ref{subsec-line_profiles} we determine that $\gtrsim$\,65\,$\%$,
$\gtrsim$\,50\,$\%$, and $\gtrsim$\,15\,$\%$ of the observed
[\OI]63\,$\upmu$m, [\OI]145\,$\upmu$m, and [\CII]158\,$\upmu$m
line luminosity arises from  the outflow. Most of the
H$_2$, CO, H$_2$O, and OH luminosity arises from the outflow and not
from \mbox{OMC-1's}  face-on PDR.
$^c$From \textit{Herschel}   \citep{Goipacs15}.
$^d$From ISO \citep{Rosenthal00}.}
\end{table}       
 
\begin{table*}[ht] 
\begin{center}
\caption{[\OI]\,63, 145\,$\upmu$m and FIR CO line luminosities toward the central emission
(a size of $\sim$0.2\,pc) of several high-mass star-forming regions.} 
\label{Table_OI_lum}  
\small
\begin{tabular}{l c c c c c c c c @{\vrule height 8pt depth 4pt width 0pt}}    

\hline\hline       
                            & $L_{\rm bol}$    & $d$       & $L_{\rm OI}$      & $L_{\rm CO}$(PACS) &  $L_{\rm CII}$ & & \\
        Source                          &   (\Lsun)$^a$    & (kpc)$^a$ & (\Lsun)           & (\Lsun)            &        (\Lsun)     & $L_{\rm OI}$\,/\,$L_{\rm bol}$ & $L_{\rm CO}$(PACS)\,/\,$L_{\rm bol}$ & $I_{63}/I_{158}$ (total line profile) \\\hline 
NGC7538-IRS1                    &    1.1E+05       &  2.7      & 11$^d$            &     1.8$^d$                 &       2.0$^d$         & 1.0E-04                                              &  1.6E-05                                                     & $\sim$5$^d$\\ 
W3-IRS5                                 &    2.1E+05       &  2.0      & 4.2$^d$           &   10.5$^d$              &   1.5$^e$         & 2.0E-05                        &  5.0E-05                                                   & $\sim$3$^e$ \\\hline 
DR21\,(G81.7+0.6)$\dagger$  &    1.3E+04       &  1.5      & $\gtrsim$10$^{b}$ &       1.2$^d$             &   2.5$^b$     & $\gtrsim$7.7E-04               &  9.2E-05                                              & $\sim$5$^b$ \\   
W28A\,(G5.89-0.39)$\dagger$ &    4.1E+04       &  1.3      & 5.7$^c$           &         3.9$^d$                 &   0.42$^c$     & 1.4E-04                                           &  9.5E-05                                              & $\sim$13$^c$ \\ 
Orion BN/KL$\dagger$        &    1.1E+05           &  0.4      & 86.5              &    64$^f$          &   5.8         & 7.9E-04                        &  5.8E-04                                                   & $\sim$14 (map), $\sim$23 (Peak\,1)\\\hline                      

\end{tabular}                                                                                            
\end{center} 
\normalsize
\vspace{-0.5cm}
\tablefoot{\mbox{[\OI]\,63, 145\,$\upmu$m}, and [\CII]\,158$\upmu$m, and CO line luminosities for the small sample of massive sources observed with \mbox{\textit{Herschel}/PACS} that exhibit clear [\OI]\,63\,$\upmu$m line emission spectra \citep{Karska14}, that is, with little contribution from foreground absorption
\citep[][]{Gerin15}.
  $\dagger$Hosts an explosive outflow. $^a$From \citet{Tak13}, \citet{Karska14}, and references therein.
$^b$Estimated
from SOFIA/FIFI-LS \citep{Karska25}. $^c$From SOFIA/GREAT velocity-resolved observations \citep{Leurini15}.
$^d$From PACS  observations 
toward the central spaxel \citep{Karska14}. $^e$From PACS  observations 
toward the central spaxel \citep{Gerin15}. $^f$From PACS maps \citep{Goipacs15}.}                 
\end{table*}      

\subsection{Mass-loss rate of the explosive outflow in Orion BN/KL}

The energy loss from the outflow-driven shock provides a measure of the 
kinetic (mechanical) luminosity of the outflow, with 
\mbox{$L_{\rm cool}\,=\,(1-f_m)\,L_{\rm kin}\,=\,(1-f_m)\,\frac{1}{2}\,\dot{M_0}\,v_{S}^2\,$}.
In this expression\footnote{This equation is written in the reverse form compared to that used in some outflow papers \citep[e.g.,][]{Maret09}, such that $L_{\rm kin} \geq L_{\rm cool}$.},
$L_{\rm cool}$ is the  total shock cooling rate 
and  \mbox{($1-f_m$)} denotes the fraction of the shock mechanical energy translated into
internal excitation.
In general, gas cooling proceeds through IR, \HI, and optical line emission,
\mbox{$L_{\rm cool} \approx L_{\rm IR} + L_{\rm HI} + L_{\rm optical}$}, 
where 
\mbox{$L_{\rm IR} = L_{\rm atoms} + L_{\rm H_2} + L_{\rm molec}$} is the contribution 
from  atomic fine-structure lines, H$_2$ lines, and other molecular lines
(CO, H$_2$O, OH, etc.).
$L_{\rm HI}$ and $L_{\rm optical}$ are only relevant in  \mbox{dissociative $J$-type} shocks \citep[][]{Lehmann20}. 
$L_{\rm HI}$ represents the  hydrogen-line cooling emission 
\mbox{(e.g., Ly$\alpha$, Ly$\beta$, and \mbox{two-photon} continuum)} and $L_{\rm optical}$ refers to ``optical'' atomic forbidden-line emission, which arises from electronic transitions  in the visible and NIR.
From the above relation$^7$,
\begin{equation}
\dot{M_0} = 1.2\cdot10^{-2}\,\frac{(1-f_m)^{-1}}{f_{\rm IR}}\,\left(\frac{L_{\rm IR}}{\Lsun}\right)\,\left(\frac{v_{S}}{{\rm{km\,s^{-1}}}}\right)^{-2}\,\,\,\Msun\,{\rm{yr^{-1}}},
\label{eq:massloss}
\end{equation}
 where $f_{\rm IR}$ is the fraction of the total cooling due to IR  emission lines.
 In general, \mbox{$f_{\rm IR}$\,$\approx$\,0.5-0.8} in \mbox{$C$-type} shocks 
 \citep[e.g.,][]{Kaufman99,Karska25} and \mbox{$f_{\rm IR}$\,$\approx$\,0.5} in \mbox{$J$-type} shocks \citep[e.g.,][]{Lehmann20}.
\mbox{Equation~(\ref{eq:massloss})} 
has clear limitations
when different types of shocks coexist,  or when part of the emitted
line luminosity arises from FUV-heated gas.
In the following analysis, we assume that \mbox{$v_S$\,$\simeq$\,30--40\,km\,s$^{-1}$} represents the velocity range of the possible shock types in  BN/KL's outflow (if more than one), whether dissociative or nondissociative.
\mbox{Table~\ref{Table_intensities}} summarizes the IR line luminosities, integrated over the entire profile, across the outflow  (dashed square in \mbox{Fig.~\ref{fig:sofia_maps_int_ratios}}).
Inserting the outflow IR line luminosity, 
\mbox{$L_{\rm IR} \simeq 320$\,\Lsun}, 
\mbox{Eq.~(\ref{eq:massloss})} yields a mass-loss rate of 
\mbox{$\dot{M}_0 \simeq (9.1 \pm 2.6) \times 10^{-3}$\,\Msun\,yr$^{-1}$}, 
using \mbox{$f_{\rm IR} = 0.5$} \citep[Fig.~3 of][]{Lehmann20} and 
\mbox{$(1-f_m)$\,=\,0.75} 
\mbox{\citep{Kaufman96}}.
Even if the derived $\dot{M}_0$ is an \mbox{approximate}  value, the estimated
mass-loss rate of BN/KL's outflow is higher than that of steady protostellar outflows.
Given a dynamical timescale of \mbox{$t_{\rm dyn}$\,$\simeq$\,500\,$\mathrm{yr}$} \mbox{\citep[e.g.,][]{Zapata09}}, the derived outflow mass is \mbox{$M = t_{\rm dyn}\,\dot{M}_0\,\simeq\,(3.3-5.9)\,\Msun$}.
This estimate agrees with,
yet may be more accurate than, mass estimates obtained from column densities computed
under the LTE assumption \citep[][]{Snell84,Peng12}. 
Still, future theoretical work should revisit the validity of \mbox{Eq.~(\ref{eq:massloss})}
in the complex scenario described in \mbox{Sect.~\ref{subsec:shock-models}}, 
considering its dependence on multiple shocks and $G_0^{\rm ext}$.

\subsection{Mass-loss rate from the putative BN outflow}

Here,  we assume that the spatially unresolved [\OI]\,63\,$\upmu$m emission detected around BN--both spatially and in LSR velocity (see \mbox{Sect.~\ref{sec:compact-outflow}})--arises from shocked gas produced by a compact outflow launched by this star.
When [\OI]\,63\,$\upmu$m is the dominant gas coolant in a \mbox{$J$-type} ``wind'' 
shock--consistent with detection of this outflow only in  
[\OI]\,63\,$\upmu$m--\citet{Hollenbach85} showed that the line luminosity scales with the outflow mass-loss rate as \mbox{$\dot{M}$\,(\Msun\,yr$^{-1}$)\,$\simeq$\,10$^{-4}$\,$L_{63}$\,(\Lsun)},
provided that the flux of material into the shock satisfies
\mbox{$n_{\rm H,0}\,v_S$\,$\lesssim$\,10$^{12}$\,cm$^{-2}$\,s$^{-1}$}.
In this regime, the particle flux is related to the [\OI]\,63\,$\upmu$m line intensity by \mbox{$n_{\rm H,0}\,v_S$\,$\simeq$\,10$^{13}$\,$I_{63}$}, 
with $n_{\rm H,0}$, $v_S$, and $I_{63}$ in cgs units.

The broad [\OI]\,63\,$\upmu$m line  emission around  BN 
is characterized by $I_{63}$\,$\simeq$\,5$\times$10$^{-2}$\,erg\,s$^{-1}$\,cm$^{-2}$\,sr$^{-1}$
and \mbox{$L_{63}\simeq0.6$\,\Lsun} in $\lesssim$\,4000\,au.
Thus, its contribution to the total $L_{63}$ of the entire map is  negligible.
Using the relations above, we obtain,
\mbox{$n_{\rm H,0}\,v_S$\,$\simeq$\,5$\times$10$^{11}$\,cm$^{-2}$\,s$^{-1}$}
and 
\mbox{$\dot{M} \simeq 6 \times 10^{-5}\,\Msun\,\mathrm{yr}^{-1}$}, which is typical of
accreting high-mass protostars \citep[e.g.,][]{Beuther02,Leurini15}.
Assuming further that the shock velocity is comparable to the  [\OI]\,63\,$\upmu$m line FWHM of the spectral component around BN, \mbox{$\Delta v_{\rm FWHM}$\,$\simeq$\,15\,km\,s$^{-1}$} \mbox{(Sect.~\ref{sec:compact-outflow})}, we derive a preshock gas density of several \mbox{$10^{5}$\,cm$^{-3}$}. 
These results indicate that the kinematics, density, and mass-loss rate associated with the compact [\OI]\,63\,$\upmu$m emission around BN are consistent with a disk-accretion–driven outflow originating from BN. Further observations of additional tracers at high angular resolution will be needed to confirm
the outflow versus wind bow shock scenario, or to constrain alternative interpretations.

\subsection{Comparison with other high-mass star-forming regions}\label{subsec:comparison}
 
 To investigate whether explosive outflows produce distinct spectroscopic features, here we compare the observed FIR [\OI], [\CII], and CO line luminosities 
of the BN/KL outflow with those measured in the central regions of other, more distant high-mass star-forming regions 
 where similar observations have been carried out \citep[e.g.,][]{Leurini15}.
In the BN/KL outflow region, the FIR atomic  line luminosity is dominated by [\OI]\,63\,$\upmu$m\footnote{\citet{Goipacs15} derived 
$L_{\rm [OI]\,63}$\,$\simeq$\,16\,\Lsun\,  from \mbox{\textit{Herschel}/PACS} maps using a nonstandard engineering observing mode 
to avoid saturation. This is a factor $\sim$5 lower than 
the  SOFIA/GREAT value.
Although we do not find an obvious explanation for this discrepancy, the
SOFIA luminosities are more consistent with KAO luminosities, 
\mbox{$L_{\rm [OI]\,63}$\,=\,50\,$\pm$\,25\,\Lsun} \citep{Werner84}.}, which is $\sim$15 times 
brighter
than the [\CII]\,158\,$\upmu$m emission. The total [\OI]\,63\,$\upmu$m and \,145\,$\upmu$m line luminosity  is remarkably high, 86.5\,\Lsun, comparable to the total H$_2$ and CO luminosity  \citep{Rosenthal00,Goipacs15}. This yields
$L_{\rm [OI]}/L_{\rm CO}\simeq 0.9$ and
$L_{\rm [OI]}/L_{\rm H_2}\simeq 0.7$. That is, about 25\% of the total IR gas cooling is through 
 [\OI]  emission lines (\mbox{Table~\ref{Table_intensities}}).
Furthermore, the contribution to $L_{\rm [OI]}$ from the broad spectral component,
neglecting the narrow spike  [\OI] emission, is  55\,\Lsun.

\mbox{Table~\ref{Table_OI_lum}} compares the FIR line luminosities toward the central
regions of several high-mass star-forming regions,
on scales of $\sim$0.2\,pc comparable to our BN/KL map. Interestingly, interferometric observations of the central regions of DR21 and \mbox{G5.89$-$0.39} reveal CO outflows exhibiting fingers and \mbox{Hubble-Lema\^itre} flow kinematics, consistent with the presence of explosive outflows \citep[e.g.,][]{Zapata13,Zapata20,Guzman24}. These other regions exhibit fractional luminosities comparable to those of BN/KL outflow,  both displaying high \mbox{$L_{\rm OI}/L_{\rm bol}$} and \mbox{$L_{\rm CO}^{\rm FIR}/L_{\rm bol}$}  ratios, $\gtrsim$10$^{-4}$. In contrast, \mbox{W3-IRS5}, a proto-Trapezium system
\mbox{\citep[e.g.,][]{Rodon08}}, and \mbox{NGC7538-IRS1}, a massive star-forming core hosting an ultracompact \HII~region, a late O--to--early B star \citep[][]{Sandell04}, exhibit lower \mbox{$L_{\rm OI}/L_{\rm bol}$} and \mbox{$L_{\rm CO}^{\rm FIR}/L_{\rm bol}$} luminosities. 
Pending more robust statistics, these high fractional luminosities may represent a distinctive signature of explosive outflows.
Still, the BN/KL outflow stands out in FIR [\OI] and CO line surface luminosities, per unit area, and the \mbox{$I_{63}/I_{158}$} intensity ratios exceed those in most regions, even when integrating over the full line profile, suggesting a denser gas environment.

\section{Summary and conclusions}

In this paper, we present velocity-resolved, sub-km\,s$^{-1}$-resolution maps
of the  [\OI]\,63 and 145\,$\upmu$m fine-structure emission lines along the wide-angle
explosive  outflow in Orion BN/KL.
We complement this data with new  [\CII]\,158\,$\upmu$m
and FIR OH  line maps, all obtained with the GREAT heterodyne receivers on board SOFIA. Our main conclusions are as follows. 

-- We separated the quiescent and the outflow (broad [\OI] line-wing) components, with the latter following the  shock-excited H$_2$ emission and the former primarily tracing the foreground, face-on PDR at the \mbox{UV-illuminated} rims of \mbox{OMC-1}.
The [\OI]\,63\,$\upmu$m line has a FWZI of $\sim$85\,km\,s$^{-1}$. The FWZI of the [\CII]\,158\,$\upmu$m line is
smaller, $\lesssim$\,50\,km\,s$^{-1}$, implying that either the highest velocity gas lacks FUV illumination or (more likely) C$^+$ is quickly  converted into other species.

-- The broad [\OI]\,145\,$\upmu$m component (\mbox{$\Delta v_{\rm FWHM} \simeq 20$\,km\,s$^{-1}$}) is narrower than the corresponding CO and OH emission (\mbox{$\Delta v_{\rm FWHM} \gtrsim 30$\,km\,s$^{-1}$}), which also exhibit larger FWZIs (\mbox{$\gtrsim$\,150\,km\,s$^{-1}$}). This suggests that the shocks producing the [\OI] emission differ either in their  nature or in exact location from those driving the highest velocity CO and OH  emission.

-- The \mbox{[\OI]\,63\,/\,145} and \mbox{[\OI]\,63\,/\,[\CII]\,158} intensity ratios attain remarkably high values in the line wings (\mbox{20–30} and \mbox{40–60}, respectively), exceeding those observed in PDRs and typical protostellar outflows.
They are consistent with dense (\mbox{$n_{\rm H} \gtrsim 10^5$–$10^6$\,cm$^{-3}$}) and warm ($T \lesssim 500$\,K) post-shock gas.

--  The fine-structure line-wing emission is consistent with originating in externally irradiated, magnetized, dissociative \mbox{$J$-type} shocks with \mbox{$v_S \simeq 30$--$40\,\rm km\,s^{-1}$} and preshock densities of a few $10^{4}\,\rm cm^{-3}$. External FUV radiation arises from surrounding fast shocks and possibly from massive (proto)stars in the region if the medium is sufficiently porous.

-- The  \mbox{[\OI]\,63 and 145\,$\upmu$m} luminosity along the outflow region is significantly high, amounting to 86.5\,\Lsun, with 55\,\Lsun\, contributed by the broad component alone. This is comparable to the  H$_2$ and CO line luminosities, yielding $L_{\mathrm{[OI]}}/L_{\mathrm{CO}} \simeq 0.9$ and $L_{\mathrm{[OI]}}/L_{\mathrm{H_2}} \simeq 0.7$. 
These values imply an outflow mass-loss rate of
 \mbox{$(9.1 \pm 2.6) \times 10^{-3}$\,\Msun\,yr$^{-1}$} and a mass of \mbox{$M$\,$\simeq$\,(3.3--5.9)\,\Msun}.

-- We report the detection of  broad [\OI]\,63\,$\upmu$m emission, slightly offset from BN but aligned with the direction of the star’s proper motion.  This \mbox{$L_{63}\simeq0.6\,\Lsun$} emission could arise from
an unresolved outflow (corresponding to a mass-loss rate of several $10^{-5}\,\Msun\,\mathrm{yr^{-1}}$) or a wind bow shock 
produced by the supersonic motion of this wind-blowing runaway B star.

 The FIR [\CII] and [\OI] line intensities are
among the brightest tracers of interstellar shocks and PDRs, probing the thermal budget and
the presence of FUV radiation. In addition, \mbox{sub-km\,s$^{-1}$-resolution} maps reveal the
 gas dynamics, allowing us to quantify the contributions of radiative
and mechanical feedback.
Unfortunately, given the cessation of SOFIA operations, very-high-spectral-resolution mapping has been carried out for only a few regions of the sky.
Therefore, a FIR heterodyne mission (airborne or space) would be the only telescope to provide this essential missing information.

\begin{acknowledgements} 
We thank our referee for a constructive report that allowed us to improve the 
presentation of our work.
This work is based on observations made with SOFIA (NASA/DLR). SOFIA was jointly operated by the Universities Space Research Association, Inc. (USRA), under NASA contract \mbox{NAS2-97001}, and the Deutsches SOFIA Institut (DSI) under the DLR contract \mbox{50 OK 0901} and \mbox{50 OK 1301} to the University of Stuttgart.
GREAT was a development by the MPI f\"ur Radioastronomie and the KOSMA/Universit\"at zu K\"oln, in cooperation with the DLR Institut für Optische Sensorsysteme. GREAT development was  financed by the participating institutes, by DLR under 
\mbox{Grants 50 OK 1102}, 1103, and 1104, and within the Collaborative Research Centre 956, funded by the Deutsche Forschungsgemeinschaft (DFG).
JRG, MGSM, and MZ thank the \mbox{Spanish} \mbox{MICIU} for funding support under grant
\mbox{PID2023-146667NB-I00}. MZ acknowledges the JdC Postdoctoral Fellowship \mbox{JDC2024-054658-I}, funded by MICIU/AEI/10.13039/501100011033 and by the ESF+.
BG and AG acknowledge the support from the Programme National ``Physique et Chimie du Milieu Interstellaire'' (PCMI) of CNRS/INSU with INC/INP funded by CEA and CNES.

\end{acknowledgements}

\bibliographystyle{aa}
\bibliography{references}

\onecolumn
\begin{appendix}\label{Sect:Appendix}

\section{Observed lines, instrument configurations, and observing modes} \label{App:Observations} 

Table~\ref{tab:spectral-lines} summarizes the instrument configurations, observing parameters, and observed FIR line properties.

\begin{table*}[bh]
      \caption[]{Observed lines, instrument configurations, and observing modes.}
         \label{tab:spectral-lines}
         \centering
         \setlength{\tabcolsep}{3.5pt} 
         \renewcommand{\arraystretch}{1.0} 
         \begin{tabular}{ccclcccclcc}
            \hline\hline
            \noalign{\smallskip}
         Cycle & Date & Config. & Transition  & $\nu_0$    & $\theta_{mb}$ & $E_{\rm up}$  &  $n_{\rm crit}$  & Instrument & Obs.Mode & Map Coverage \\
               &      &        &      \textbf{($\lambda_0$)}      &                &            &               &                          &            &           & $\Delta$RA$\times\Delta$Dec\\        
               &      &        &                                              &    (GHz)   &    ($''$)     &  (K)      & (cm$^{-3}$)  &            &           & ($''\,\times\,''$)\\        

   \noalign{\smallskip}
   \hline
   \noalign{\smallskip}            
9 & 23.11.21 & LFA-H & \Cp    & 1900.53690 (L) & 14.8 &  91  & 6$\times$10$^3$ &  upGREAT     & TP OTF   & 144$\times$123 by 3$''$\\
9 & 19.11.21 & LFA-H &  (\textbf{157.7\,$\upmu$m})      &                &      &      &                 &              &          & \\
8 & 09.03.21 & LFA-H &        & 1900.53690 (U) &      &      &                 &              &          & \\
8 & 24.02.21 & LFA-H &        &    &      &      &                 &              &          & \\
  \noalign{\smallskip}
  \hline
  \noalign{\smallskip}
4 & 08.11.16 & LFA-V & \12OH & 1834.74615 (U) & 14.6 & 268  & > 10$^9$  & upGREAT & Ch OTF & 108$\times$84 by 6$''$\\
4 & 09.11.16 & LFA-V &  (\textbf{163.4\,$\upmu$m})   &             &      &      &                  &         &        &                    \\
  \noalign{\smallskip}
  \hline
  \noalign{\smallskip}
8 & 11.02.21 & 4G-4 & \32OH  & 2514.32943 (L) & 11.3 &  120 & 5$\times$10$^8$  & 4GREAT  & Ch Ras & 24$\times$24 by 6$''$\\
1 & 31.01.14 & M    &  (\textbf{119.2\,$\upmu$m})      & 2514.32943 (U) &      &      &                  & GREAT   & Ch Ras & 40$\times$40 by 10$''$\\
  \noalign{\smallskip}
  \hline
  \noalign{\smallskip}
8 & 09.03.21 & LFA-V  & \PzeroOi & 2060.06886 (U) & 12.9 & 327 & 6$\times$10$^6$ &  upGREAT  & TP OTF & 144$\times$123 by 3$''$\\
8 & 24.02.21 & LFA-V  &  (\textbf{145.6\,$\upmu$m})        &                &      &     &                 &           &        & \\
   \noalign{\smallskip}
  \hline
  \noalign{\smallskip}
9 & 23.11.21 & HFA-V   & \PoneOi  & 4744.77749 (L) & 6.3 &  228 & 5$\times$10$^5$ & upGREAT   & TP OTF & 144$\times$123 by 3$''$\\
9 & 19.11.21 & HFA-V   &  (\textbf{63.2\,$\upmu$m})        &                &     &      &                 &           &        & \\
8 & 09.03.21 & HFA-V   &          &                &     &      &                 &           &        & \\
8 & 24.02.21 & HFA-V   &          &                &     &      &                 &           &        & \\
  \noalign{\smallskip}
  \hline 
 \end{tabular}     
     
         \tablefoot{For each of the transitions studied in this work, here we quote the SOFIA observing cycle, the UT date of the observation and the instrument configuration. Rest frequencies (in parenthesis, the sideband used for the LO tuning), upper level energies above ground and characteristic critical densities are adopted and calculated from the Leiden Atomic and Molecular Database, LAMDA \citep{Schoier2005}. During observing cycle 1 single-pixel GREAT channels were operated \citep{Heyminck2012}; since cycle~4, the dual-color upGREAT array spectrometers were operated, which made it possible to record up to three different frequencies simultaneously (for details, see \citealt{Risacher2016a}, \citealt{Risacher2016b}, and \citealt{Risacher2018}). In 2021 the multi-color 4GREAT was added to the suite of GREAT flight configurations \citep{Duran2021}. Half-power beam width are quoted from these publications.  \\
         The observing modes are indicated: total power (TP) or chopped (ch) on-the-fly (OTF: data is acquired while slewing the telescope), or raster mode (the telescope stays pointed while integrating). The size of a given map is quoted for the central array pixel only; the step between dumps (along the scanning direction) and between adjacent OTF slews is given or the grid used for pointed raster observations.  
         }
\end{table*}

\clearpage

\section{Complementary observational figures} \label{App:channel_maps}

This appendix presents additional figures that aim to better understand
the velocity structure of BN/KL outflow.

\subsection{Velocity channel maps} 

\begin{figure*}[h]
\centering   
\includegraphics[scale=0.6,angle=0]{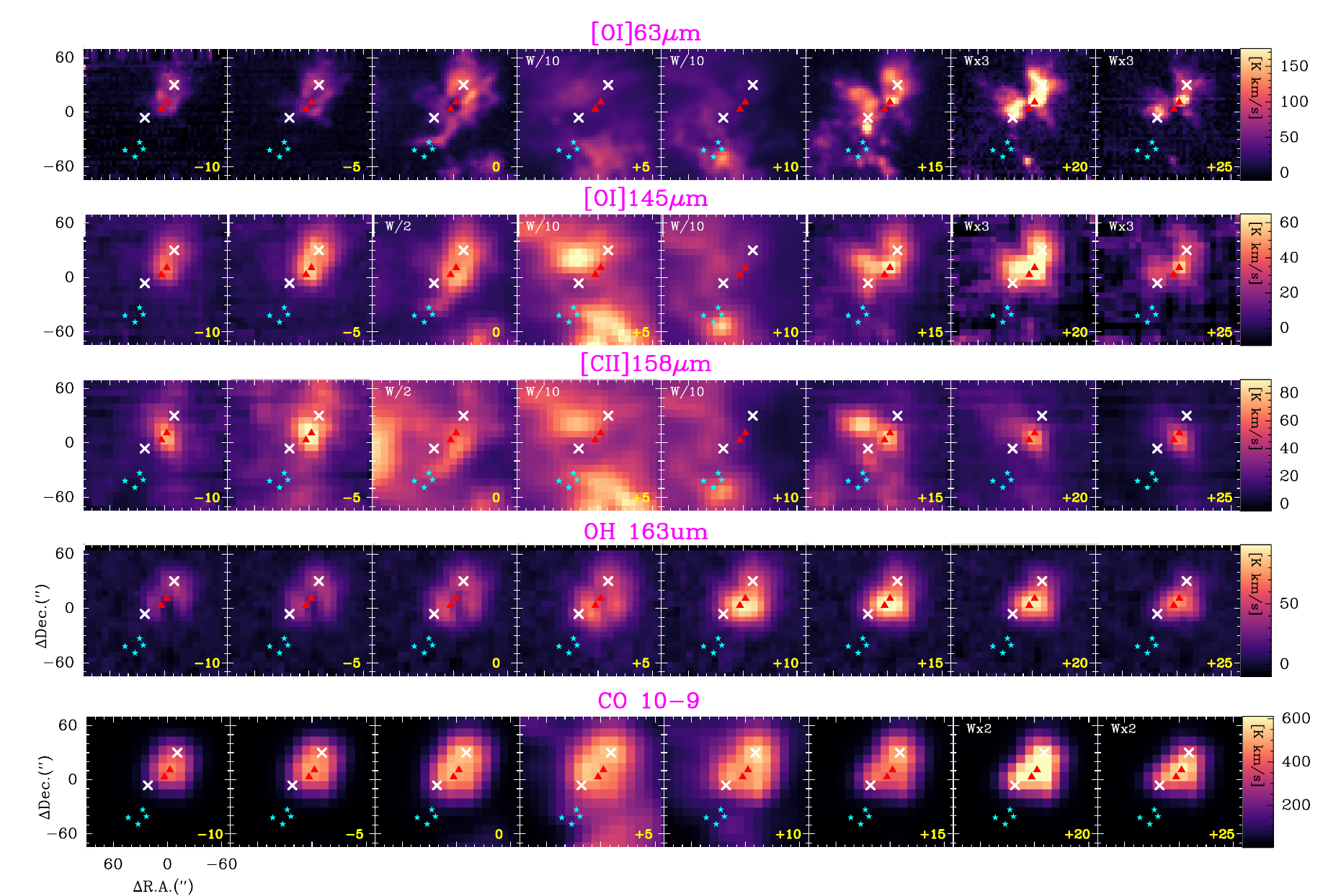}
\caption{Velocity channel maps at their native angular resolution and in K\,km\,s$^{-1}$ intensity units, from 
\mbox{$v_{\rm LSR}$\,=\,$-$10 to $+$25\,km\,s$^{-1}$} in bins of 5\,km\,s$^{-1}$ 
\mbox{[$v$, $v+5$]}, with $v$ indicated in the bottom-right corner of each panel.
The map at \mbox{$v$\,=\,$+$5\,km\,s$^{-1}$} shows the emission at the systemic velocity of the
quiescent gas in \mbox{OMC-1}, \mbox{$v_{\rm LSR,0}$\,$\simeq$\,8--9\,km\,s$^{-1}$}, dominated by extended emission from
a face-on PDR.
The main reference positions discussed in the text are illustrated with symbols
(see Fig.~\ref{fig:sofia_maps_int}).}  
\label{fig:channels}
\end{figure*}

Figure~\ref{fig:channels} shows velocity-channel maps (in K\,km\,s$^{-1}$ units) of the 
FIR lines observed with SOFIA and the CO\,$J$\,=\,10--9 line observed with 
\textit{Herschel}/HIFI
\citep{Goico19}.

\clearpage

\begin{figure*}[h]
\centering   
\includegraphics[scale=0.7,angle=0]{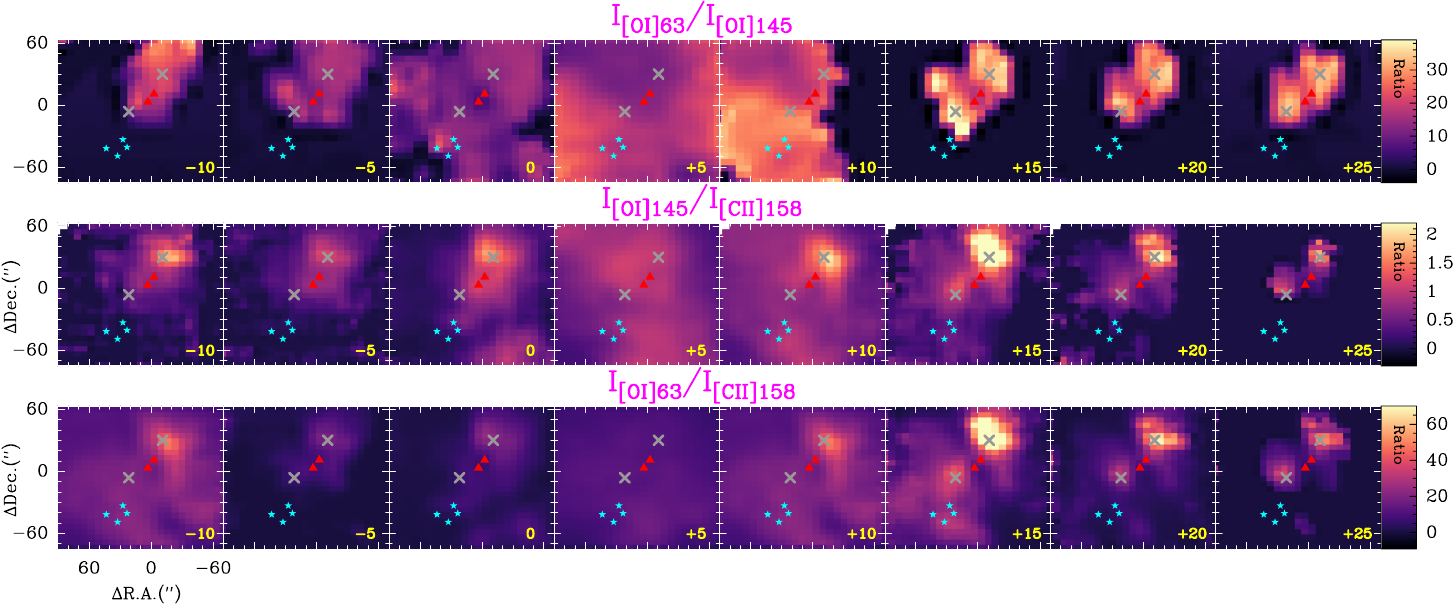}
\caption{Velocity-resolved line intensity ratio channel maps 
(same velocity bins as in \mbox{Fig.~\ref{fig:channels}})
convolved at a common
angular resolution of 15$''$, with the integrated
line intensities in erg\,s$^{-1}$\,cm$^{-2}$\,sr$^{-1}$. The map at \mbox{$v$\,=\,$+$5\,km\,s$^{-1}$} shows the emission at the systemic velocity of the
quiescent gas.}  
\label{fig:channels_ratios}
\end{figure*}

Figure~\ref{fig:channels_ratios} shows channel maps of line intensity ratios for the
different FIR atomic lines (with intensities in cgs units).

\subsection{Line profiles toward the FIR continuum peak}

\begin{figure*}[h]
\centering   
\includegraphics[scale=0.53,angle=0]{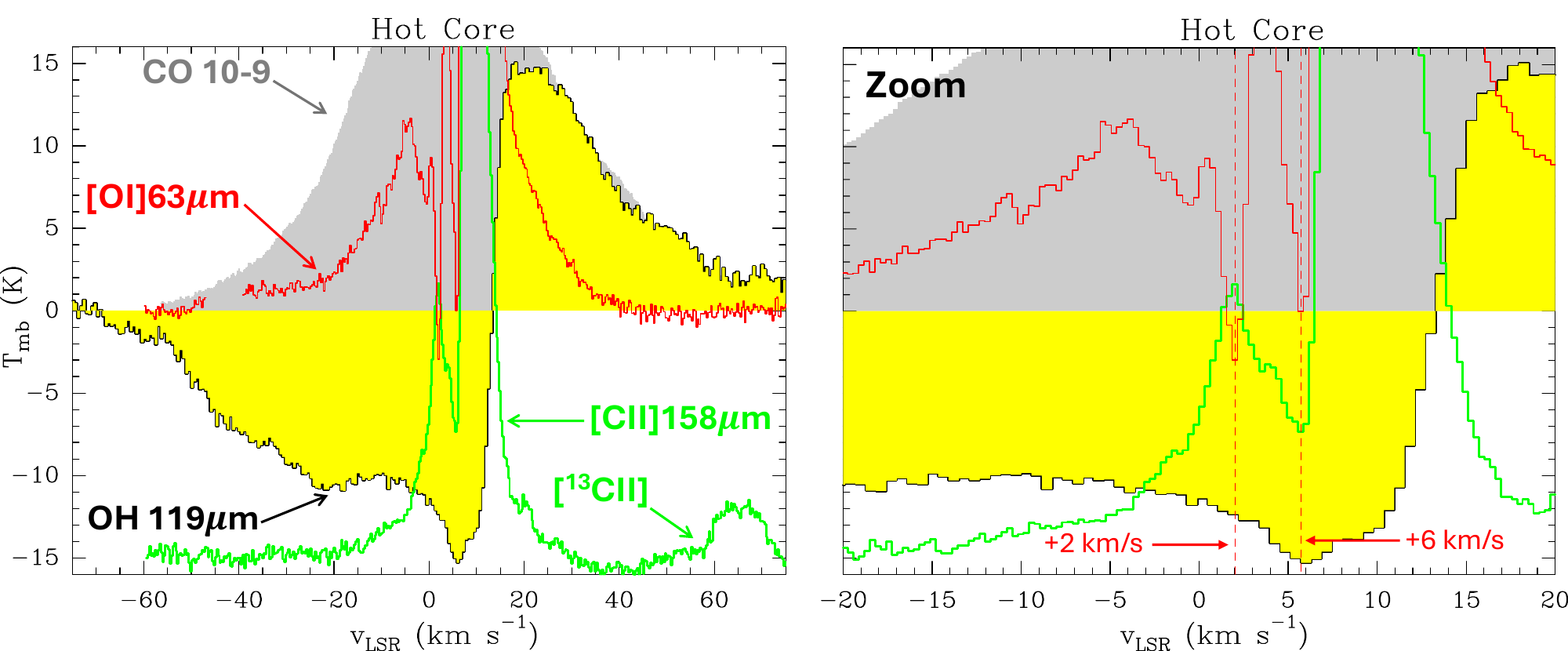}
\caption{OH\,119\,$\upmu$m \mbox{P-Cygni} profile toward the hot core (FIR continuum peak), compared to other line profiles. The OH\,119\,$\upmu$m absorption dip agrees with [\OI]\,63\,$\upmu$m absorption at $+$6\,km\,s$^{-1}$ (hot core velocities). The [\CII]\,158\,$\upmu$m emission
and [\OI]\,63\,$\upmu$m absorption at $+$2\,km\,s$^{-1}$ corresponds to a gas structure in the foreground.}  
\label{fig:ZOOM_ALL_OH}
\end{figure*}

Figure~\ref{fig:ZOOM_ALL_OH} (right panel) shows a zoom on the low-intensity features of several FIR lines toward the FIR continuum peak position (the hot core region), with
the main [\OI]\,63\,$\upmu$m absorption and [\CII]\,158\,$\upmu$m emission features labelled. 

\clearpage

\section{Nonlocal NLTE [\OI] radiative transfer models}
\label{App:MTC_models}

In our [\OI] excitation and radiative transfer models \citep{Goico06b,Goico09}, we adopted
the fine-structure \mbox{O--H$_2$} and \mbox{O--H} collisional rate coefficients calculated by \citet{Lique18}
in the 10--1000\,K temperature range. As observed, the FIR [\OI] fine-structure line emission in the outflow arises from molecular gas with some atomic H \citep{vanderWerf13}. Hence, we define the gas density as \mbox{$n_{\rm H} = n(\mathrm{H}) + 2n(\mathrm{H}_2)$}, and adopted \mbox{$n(\mathrm{H}_2)/n(\mathrm{H}) \simeq 4$} in the models. We verified that this choice does not affect our main conclusions and results. 
We ran spherical models with uniform gas densities ($n_{\rm H} = 10^4$–10$^8$\,cm$^{-3}$), temperatures ($T_{\rm k} = 200$ and 500\,K, based on the observed \mbox{$T_{\rm mb}$([\OI]\,63\,$\upmu$m)} line peaks), and a fixed velocity dispersion, \mbox{$\sigma_{\rm turb} = 8.5$\,km\,s$^{-1}$} (i.e., $\Delta v_{\rm turb,\,FWHM} \simeq 20$\,km\,s$^{-1}$), matching the observed [\OI]\,145\,$\upmu$m broad component toward Peak~1.

\begin{figure*}[b]
\centering
\includegraphics[scale=0.54,angle=0]{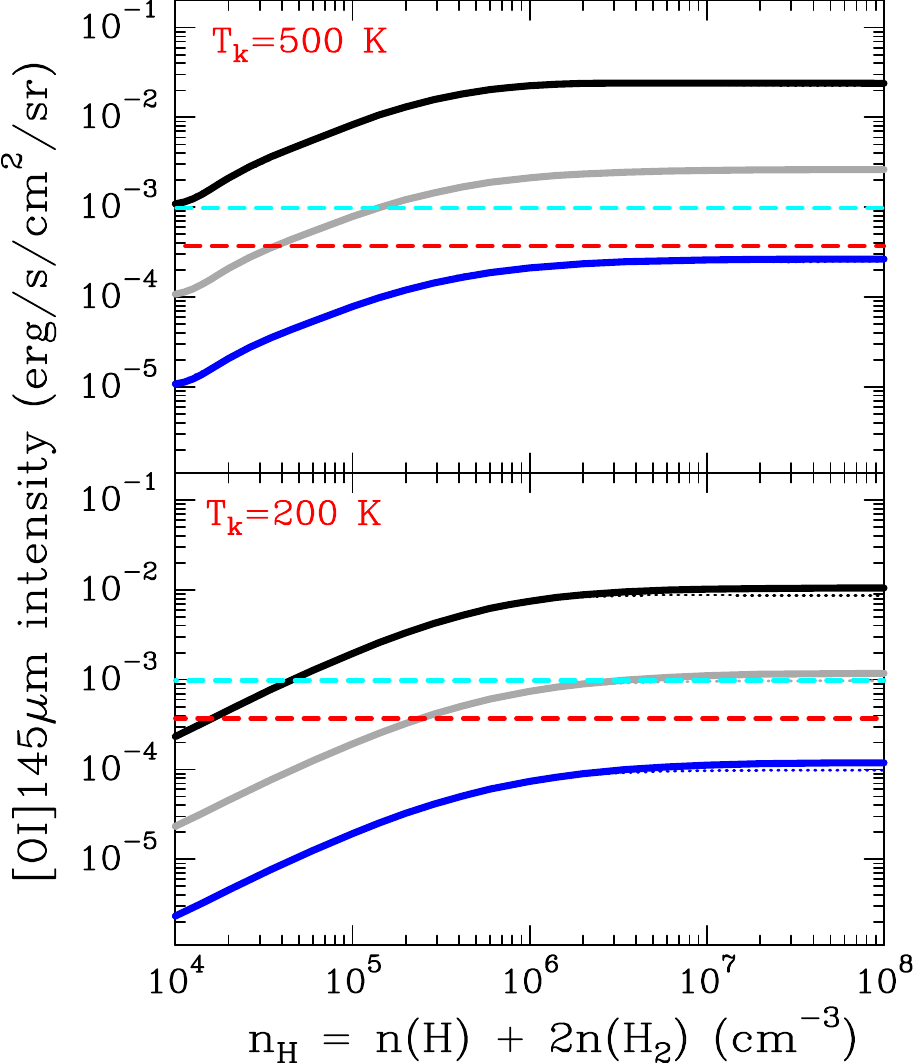} \hspace{1.5cm}
\includegraphics[scale=0.425,angle=0]{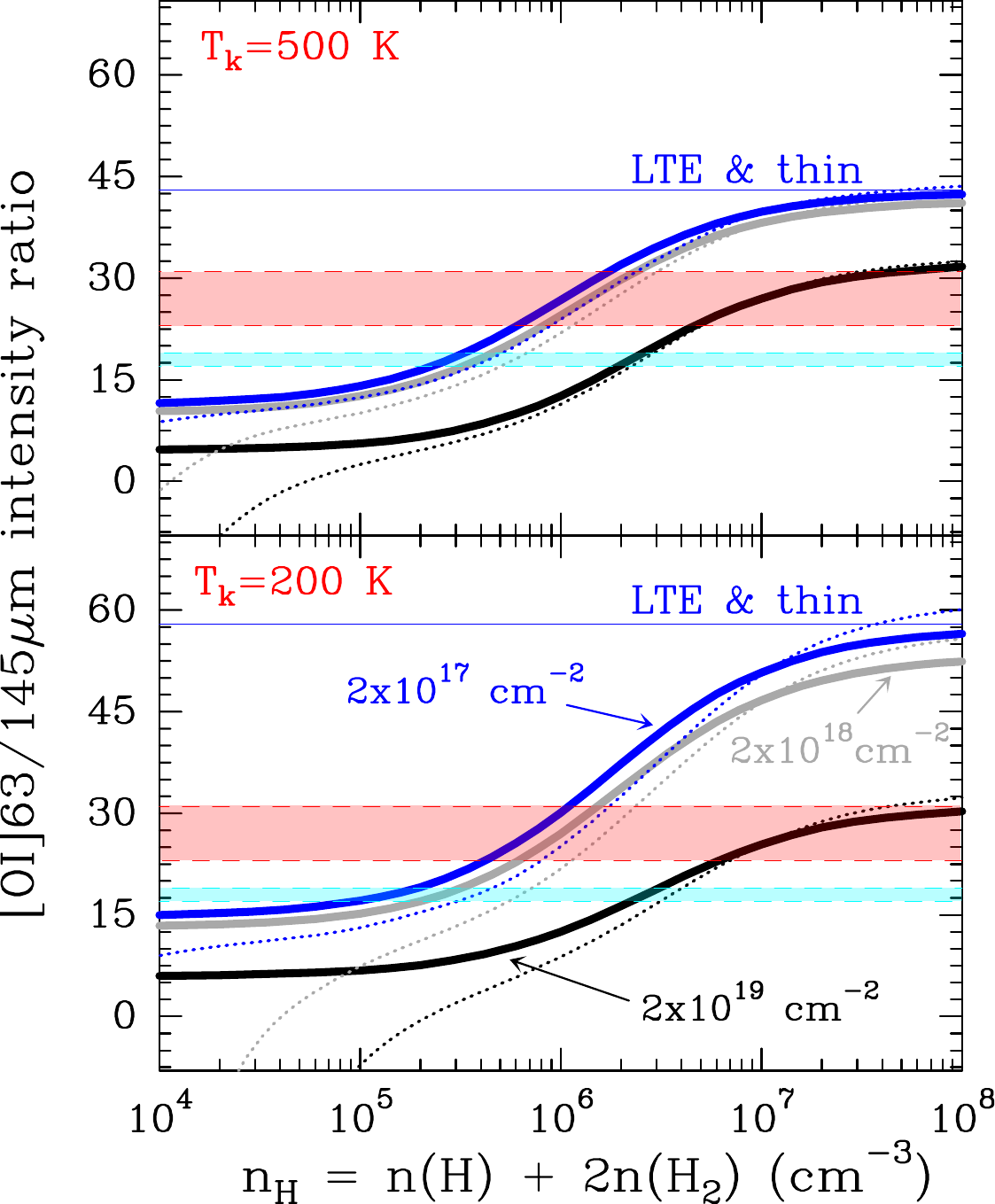} 
\caption{Nonlocal and NLTE [\OI] excitation and radiative transfer
models: Predicted [\OI]\,145\,$\upmu$m line intensity (\textit{left}) and
[\OI]\,63/145\,$\upmu$m intensity ratio in cgs units (\textit{right})
 as a function of total gas density. The black, grey, and blue curves represent
 different column densities of atomic oxygen. 
 Dotted curves represent models with external FIR illumination (see text).
  In the left panels, the red and cyan dashed lines show the [\OI]145\,$\upmu$m intensities measured toward Peak~1 in the red- and blueshifted wings, respectively.
 In the right panels, the red and cyan dashed colored areas show the intensity ratios
 (with errors) measured toward Peak~1 in the red- and blueshifted wings, respectively (see Table~\ref{Table_intensities_wings}).}
\label{fig:oi_145_int}
\end{figure*}

Figure~\ref{fig:oi_145_int} shows the results of the grid, displaying $I_{145}$ as a function of $n_{\rm H}$ (left panels) for different atomic oxygen column densities: 2\,$\times$\,10$^{17}$, 2\,$\times$\,10$^{18}$, and 2\,$\times$\,10$^{19}$\,cm$^{-2}$, and gas temperatures. 
The right panels shows the resulting $I_{63}$/$I_{145}$ line intensity ratio as a function of $n_{\rm H}$. Continuos curves refer to models without external FIR illumination.
The red and cyan colored areas show
the  $I_{63}$/$I_{145}$ ratios
measured toward Peak 1 in the red- and blueshifted wings, respectively
(extracted from Table~\ref{Table_intensities_wings}). 
 These ratios are consistent with the presence of
dense ($n_{\rm H}$\,$\simeq$\,several 10$^5$ to 10$^6$\,cm$^{-3}$) and warm ($T$\,$\lesssim$\,500\,K) \mbox{post-shock} gas,
with  \mbox{$N$(O)\,$\simeq$ a few 10$^{18}$\,cm$^{-2}$}.
 
\begin{figure*}
\centering
\includegraphics[scale=0.54,angle=0]{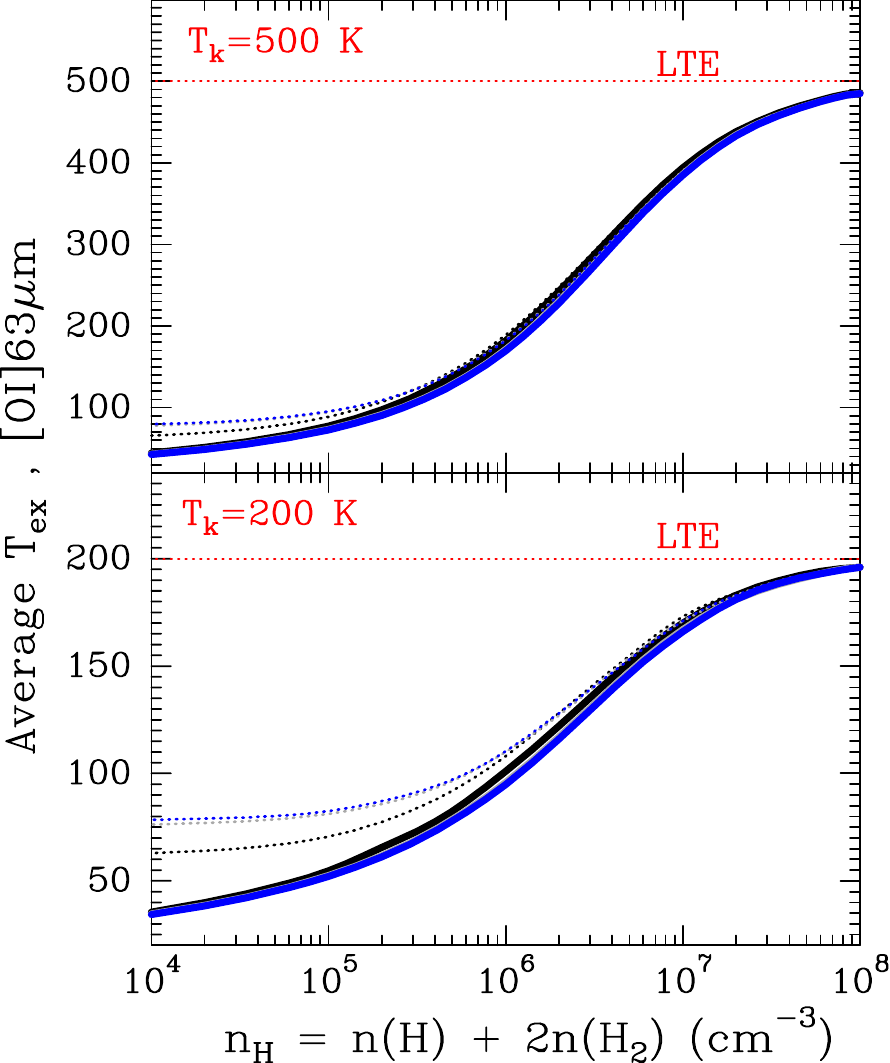} \hspace{1.5cm}
\includegraphics[scale=0.54,angle=0]{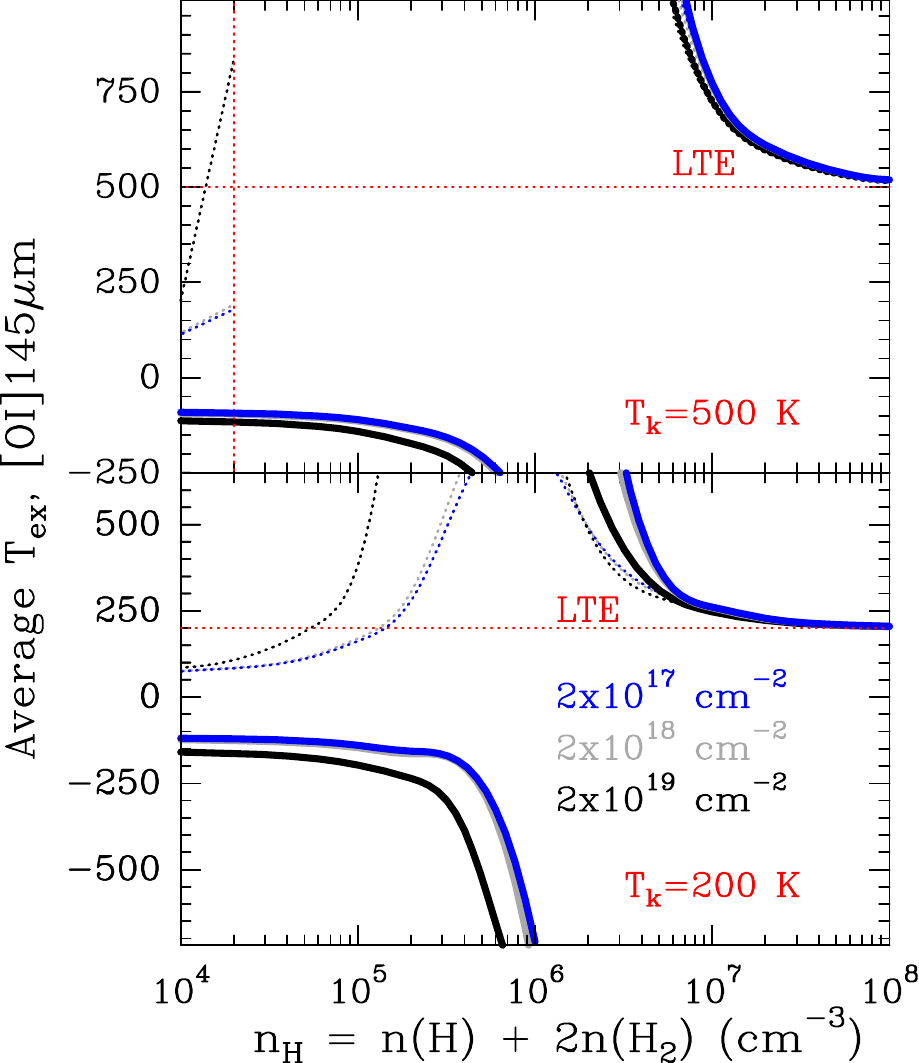} 
\caption{Average excitation temperature of the [\OI]\,63\,$\upmu$m and
[\OI]\,145\,$\upmu$m lines as a function of gas density for two gas temperatures,
200 and 500\,K. The black, grey, and blue curves represent
 different column densities of atomic oxygen, indicated in the right panel.
  Dotted curves represent models with external FIR illumination (see text).
  The increase in the [\OI]\,63\,$\upmu$m excitation temperature at low
  densities reflects
  the role of FIR pumping by a modified black body at \mbox{$T_{\rm d}$\,=\,100\,K}.
  This effect reduces the population
  inversion of the  transition at 145$\upmu$m at low densities.
  }
\label{fig:oi_Tex}
\end{figure*}

In these plots, the opacity of the [\OI]\,63\,$\upmu$m line increases
as $N$(O) increases, from optically thin to thick emission. That is,
from high to low densities--from right to left in the figures--as collisional excitation becomes less important and the
low energy levels become more populated. This is reflected in very sub-thermal [\OI]\,63\,$\upmu$m excitation (\mbox{$T_{\rm ex,\,63\mu m} \ll T_{\rm k}$}) when \mbox{$n_{\rm H} \ll n_{\rm cr,\,63\,\mu m}$} (\mbox{Fig.~\ref{fig:oi_Tex}}, left panels).
\mbox{Figure~\ref{fig:oi_Tex}} shows the average $T_{\rm ex,\,63\,\mu m}$ and $T_{\rm ex,\,145\,\mu m}$ in the models, but we note that due to radiative transfer effects—such as line trapping—the local $T_{\rm ex}$ values vary with position, even under uniform physical conditions. The excitation of the [\OI]\,145\,$\upmu$m line is more complex and is governed by the spontaneous radiative decay of the upper level ($^3$P$_0$), which is roughly five times lower than that of the lower level ($^3$P$_1$). In addition, the collisional de-excitation rate coefficient for \mbox{O–H$_2$} collisions of the \mbox{$^3$P$_0$--$^3$P$_1$} (145\,$\upmu$m) transition is significantly smaller than that of the \mbox{$^3$P$_1$--$^3$P$_2$} (63\,$\upmu$m) transition \citep[e.g.,][]{Monteiro87,Lique18}.  These properties lead to [\OI]\,145\,$\upmu$m population inversions
 \mbox{($T_{\rm ex,\,145\mu m}$\,$<$\,0\,K)}
at low densities (\mbox{$n_{\rm H} < n_{\rm cr,\,145\,\mu m}$}) and suprathermal emission \mbox{($T_{\rm ex,\,145\,\mu m}$\,$>$\,$T_{\rm k}$)}
at high densities   \citep[previously discussed by][]{Liseau06,Elitzur06,Goico09,Goldsmith19}.
These features are evident in the right panels of Fig.~\ref{fig:oi_Tex}.

To investigate the role of FIR pumping in the [\OI] excitation, 
the dotted curves in Figs.~\ref{fig:oi_Tex} and \ref{fig:oi_145_int} represent the same models but adding external FIR continuum illumination.
We model the continuum background as the sum of the cosmic microwave background
and a modified blackbody, \mbox{$\eta\,B$($T_{\rm d}$)(1\,$-$\,exp($-\tau_{\rm d,\,\lambda})$)}, at a dust color temperature of \mbox{$T_{\rm d}$\,=\,100\,K}.
In this expression, $\eta$ is the filling factor of the background FIR continuum emission, which we set to 0.5. The
dust continuum emission is assumed to be optically thick at 100\,$\upmu$m, with
\mbox{$\tau_{\rm d,\,\lambda}$\,=\,7\,[100/$\lambda$($\upmu$m)]$^2$}, and represents the typical FIR dust continuum emission close to the hot core region \citep[see][]{Goipacs15}.
The main role of this FIR continuum is to produce [\OI]\,63\,$\upmu$m
line absorption (when \mbox{$T_{\rm ex}$\,$<$\,$T_{\rm cont}$}) at low gas densities, $n_{\rm H}$ below a few 10$^4$\,cm$^{-3}$,
as the [\OI]\,63\,$\upmu$m line becomes optically thick, for 
\mbox{$N$(O) $\gtrsim$  10$^{18}$\,cm$^{-2}$}.

In general, FIR pumping increases the population of the $^3$P$_1$ level (in this case increasing the excitation temperature of the [\OI]\,63\,$\upmu$m line; see dotted curves in Fig.~\ref{fig:oi_Tex}, left), thereby reducing the population inversion of the $^3$P$_0$–$^3$P$_1$ transition at [\OI]\,145\,$\upmu$m. At low to moderate gas densities, the [\OI]\,63\,$\upmu$m line intensity decreases, whereas the intensity of the [\OI]\,145\,$\upmu$m remains nearly unchanged. Consequently, the $I_{63}/I_{145}$ intensity ratio is slightly lower (dotted curves in \mbox{Fig.~\ref{fig:oi_145_int}}, right) than in the nonilluminated case. Hence, the inferred post-shock gas density may be slightly higher.

At high densities, however, the main effect of FIR pumping is to slightly reduce $T_{\rm ex,\,145\mu m}$, although it remains suprathermal (\mbox{Fig.~\ref{fig:oi_Tex}}, right), leading to a slightly higher $I_{63}/I_{145}$ intensity ratio (\mbox{Fig.~\ref{fig:oi_145_int}}, right).

\clearpage

\section{Line intensity tables}
\label{App:Line_intensities}

In this appendix we tabulate the FIR line intensities toward the Trapezium position and Peak~1, as well as the average line intensities over the entire outflow region.

\begin{table*}[!h] 
\begin{center}
\caption{Results of Gaussian fits to FIR lines toward the narrow [\OI]\,63\,$\upmu$m line intensity peak near the Trapezium, at an offset of $(+20'', -50'')$.}  
\label{Table_int_Trapezium}  
\normalsize
\begin{tabular}{c c c c c c @{\vrule height 10pt depth 5pt width 0pt}}    
\hline\hline       
Line                                                 &      $W$             &  $I$ & $v_{\rm LSR,\,peak}$                         & $\Delta v$     & $T_{\rm mb,\,peak}$\\
                                                     &   (K\,km\,s$^{-1}$)  & (erg\,s$^{-1}$\,cm$^{-2}$\,sr$^{-1}$)  &  (km\,s$^{-1}$)      & (km\,s$^{-1}$) &     (K)\\\hline
$[$\OI$]$\,63\,$\upmu$m          &    1496.9$\pm$1.0    &       1.638E-01                                                & 11.1$\pm$0.1       &   5.4$\pm$0.1    &    261.7\\
$[$\OI$]$\,145\,$\upmu$m         &     878.3$\pm$1.8    &       7.868E-03                                            & 10.5$\pm$0.1       &   5.0$\pm$0.1    &    166.0\\
$[$\CII$]$\,158\,$\upmu$m        &    1259.4$\pm$5.1    &       8.886E-03                                                & 10.4$\pm$0.1       &   5.5$\pm$0.1    &    215.2\\
CO $J$\,=\,10--9$^{\dagger}$     &    348.0$\pm$0.9     &       5.451E-04                                                & 9.9$\pm$0.1        &   4.5$\pm$0.1     &     72.5\\
OH\,163\,$\upmu$m                &    11.8$\pm$3.0      &       7.467E-05                                                & 10.8$\pm$0.4       &   4.0$\pm$1.8    &      2.8\\\hline
\end{tabular}                                                                                             
\end{center} 
\normalsize
\vspace{-0.5cm}
\tablefoot{Line intensities in a beam of 15$''$, from maps convolved to a common angular resolution of 15$''$,
except for \mbox{CO 10--9}.\\ $\dagger$From \textit{Herschel}/HIFI observations of \citet{Goico19}. }                
\end{table*}      

\begin{table*}[!h] 
\begin{center} 
\caption{Observed FIR line intensities toward Peak 1 as a function of velocity, binned in intervals of 5\,km\,s$^{-1}$ (see text and Fig.~\ref{fig:sofia_maps_int_ratios}).} 
\label{Table_intensities_vPeak1}  
\normalsize
\begin{tabular}{c c c c c @{\vrule height 10pt depth 5pt width 0pt}}    
\hline\hline       
 [$v_{\rm LSR}$, $v_{\rm LSR}$\,+\,5] & [\OI]\,63\,$\upmu$m                    &          [\OI]\,145\,$\upmu$m           &  [\CII]\,158\,$\upmu$m                 &  OH\,163\,$\upmu$m \\
  (km\,s$^{-1}$)    &  ($\times$10$^{-3}$ erg\,s$^{-1}$\,cm$^{-2}$\,sr$^{-1}$) &    &   &  \hspace{-5cm} ($\times$10$^{-4}$ erg\,s$^{-1}$\,cm$^{-2}$\,sr$^{-1}$) \\\hline
 $[-20,-15]$                                              &  3.0$\pm$0.1                                        &   1.6$\pm$0.2                                   &   0.8$\pm$0.1     & 1.7$\pm$0.2  \\
 $[-15,-10]$                                              &  4.6$\pm$0.1                                            &   2.3$\pm$0.2                                           &   1.1$\pm$0.1      & 1.8$\pm$0.2  \\
 $[-10,-5]$                                           &  6.7$\pm$0.1                                &   3.1$\pm$0.2                                           &   1.7$\pm$0.1      & 2.4$\pm$0.2  \\
  $[-5,0]$                                                &  7.4$\pm$0.1                                    &   4.6$\pm$0.2                                           &   3.9$\pm$0.1      & 2.6$\pm$0.2  \\
  $[0,+5]$                                                &  12$\pm$1                               &   8.7$\pm$0.2                                           &   5.5$\pm$0.1      & 3.0$\pm$0.2  \\
  $[+5,+10]$                                              &  40$\pm$1                                       &   36$\pm$1                                          &   35$\pm$1          & 3.5$\pm$0.2 \\
 $[+10,+15]$                                              &  28$\pm$1                              &   12$\pm$1                                           &   6.4$\pm$0.1          & 3.5$\pm$0.2  \\
 $[+15,+20]$                                              &  12$\pm$1                                  &   3.9$\pm$0.2                                        &   1.2$\pm$0.1          & 3.3$\pm$0.2  \\
 $[+20,+25]$                                              &  5.5$\pm$0.1                                        &   1.8$\pm$0.2                                                    &   0.8$\pm$0.1         & 2.9$\pm$0.2 \\
 $[+25,+30]$                                              &  2.8$\pm$0.1                                        &   1.0$\pm$0.2                                                    &   0.5$\pm$0.1         & 2.1$\pm$0.2 \\\hline

\end{tabular}                                                                                             
\end{center} 
\normalsize
\vspace{-0.5cm}
\tablefoot{Line intensities from maps convolved to a common angular resolution of 15$''$.}                
\end{table*}      

\begin{table*}[!h] 
\begin{center}
\caption{Observed FIR line intensities in Orion's BN/KL outflow as a function of velocity, binned in intervals of 5\,km\,s$^{-1}$, and averaged over an area of 
\mbox{100\,$''\times$80\,$''$} (see text and Fig.~\ref{fig:sofia_maps_int_ratios}).} 
\label{Table_intensities_voutflow}  
\normalsize
\begin{tabular}{c c c c c @{\vrule height 10pt depth 5pt width 0pt}}    
\hline\hline       
 [$v_{\rm LSR}$, $v_{\rm LSR}$\,+\,5] & [\OI]\,63\,$\upmu$m                    &          [\OI]\,145\,$\upmu$m           &  [\CII]\,158\,$\upmu$m                 &  OH\,163\,$\upmu$m \\
  (km\,s$^{-1}$)                      &  ($\times$10$^{-3}$ erg\,s$^{-1}$\,cm$^{-2}$\,sr$^{-1}$) &   &   &  \hspace{-5cm} ($\times$10$^{-4}$ erg\,s$^{-1}$\,cm$^{-2}$\,sr$^{-1}$) \\\hline
 $[-20,-15]$                                              &   0.5$\pm$0.1                                    & 0.1$\pm$0.1                        &  0.2$\pm$0.1                                      & 0.3$\pm$0.1  \\
 $[-15,-10]$                                              &   0.9$\pm$0.1                                    & 0.3$\pm$0.1                        &  0.3$\pm$0.1                                      & 0.5$\pm$0.1  \\
 $[-10,-5]$                                           &   1.5$\pm$0.1                          & 0.6$\pm$0.1                     &  0.8$\pm$0.1                                            & 0.6$\pm$0.1  \\
  $[-5,0]$                                                &   1.9$\pm$0.1                                & 1.2$\pm$0.1                    &  1.9$\pm$0.1                                           & 0.8$\pm$0.1  \\
  $[0,+5]$                                                &   4.2$\pm$0.1                                & 3.2$\pm$0.1                    &  4.6$\pm$0.1                                           & 1.1$\pm$0.1  \\
  $[+5,+10]$                                              &   35$\pm$1                                     & 28$\pm$1                            &  31$\pm$1                                           & 1.6$\pm$0.1  \\
 $[+10,+15]$                                              &   25$\pm$1                                 & 11$\pm$1                                &  14$\pm$1                                       & 1.8$\pm$0.1  \\
 $[+15,+20]$                                              &   5.0$\pm$0.1                                    & 1.6$\pm$0.1                        &  1.3$\pm$0.1                                  & 1.6$\pm$0.1  \\
 $[+20,+25]$                                              &   1.9$\pm$0.1                                    & 0.5$\pm$0.1                        &  0.5$\pm$0.1                                  & 1.2$\pm$0.1  \\\hline
\end{tabular}                                                                                             
\end{center} 
\normalsize
\vspace{-0.5cm}
\tablefoot{Line intensities in a beam of 15$''$, from maps convolved to a common angular resolution of 15$''$.}                
\end{table*}      

\end{appendix}

\end{document}